\documentclass[pdflatex,sn-mathphys-num]{sn-jnl}
\usepackage{amssymb}
\usepackage{amsmath}
\usepackage{textcomp}
\usepackage{gensymb}
\usepackage{subcaption}
\usepackage{subfiles}
\usepackage{siunitx}
\usepackage{placeins}
\usepackage{url}
\usepackage{etoolbox}
\usepackage{xcolor}
\usepackage{booktabs}
\usepackage[version=4]{mhchem}
\usepackage{multirow}
\usepackage{graphicx}
\usepackage{adjustbox}
\usepackage{xr-hyper}       
\usepackage{hyperref}

\begin{document}

\title[Article Title]{A Consistent Interface Reconstruction and Coupling Method for Multiphysics Simulations}

\author[1]{\fnm{Ethan} \sur{Huff}}\email{ethan.huff@uky.edu}
\author*[1]{\fnm{Savio J.} \sur{Poovathingal}}\email{saviopoovathingal@uky.edu}

\affil[1]{\orgdiv{Department of Mechanical and Aerospace Engineering}, \orgname{University of Kentucky}, \orgaddress{\street{287 Ralph G. Anderson Building}, \city{Lexington}, \postcode{40506}, \state{Kentucky}, \country{USA}}}

\abstract{ Accurate representation of interfaces and flux exchange is vital for coupled multiphysics simulations across a broad range of applications. Currently, coupling approaches are limited by the underlying discretization or to specific physical problems, restricting their generality. To remove these constraints, a consistent interface reconstruction and coupling method has been developed to bridge multiphysical computational domains. The proposed numerical framework contains two complementary steps. The first step reconstructs a continuous bounding surface from discretized spatial data using a weighted interpolation and marching-grid approach that preserves geometric fidelity across a wide range of resolutions. The second step consists of a conservative flux mapping algorithm that projects surface quantities such as aerodynamic loads, heat fluxes, or mass transfer onto nearby discrete elements while maintaining global conservation of flux quantities. This new numerical formulation allows for integration of the various partitioned domains and ensures consistent data transfer between them. The framework is tested on various geometries, and surface containment errors below 2.5\% and flux-transfer errors below 1\% are obtained. Transient simulations of uniform surface recession further confirm accuracy in coupled evolution, with predicted volume loss agreeing with analytical solutions to within 1\%. The proposed method establishes a general and extensible framework for consistent interface reconstruction and coupling in multiphysics environments, providing a pathway toward unified treatment of discrete-continuum interactions for a wide range of systems.
}




\maketitle


\section*{Article Highlights}
\begin{itemize}
\item A general method for connecting grid-based solids with surface-based simulations.
\item Preserves geometric evolution and surface fluxes, enabling consistent data exchange across physical models.
\item Accurately captures shape evolution and material loss in coupled simulations.
\end{itemize}

\clearpage

\section{Introduction}
\label{sec_intro}


Real-world engineering systems—such as aerospace vehicles, energy conversion devices, and advanced manufacturing processes—often involve tightly coupled interactions between multiple physical phenomena~\cite{dede2014multiphysics,lani2013coolfluid,keyes2013multiphysics}. Capturing these interactions through multiphysics simulations is essential for predicting system performance, guiding design optimization, and reducing experimental cost. Such simulations integrate distinct numerical solvers to represent interacting physical processes that may operate on different spatial and temporal scales. However, achieving accurate coupling between these solvers remains a major challenge, particularly when each method employs its own discretization and geometric representation~\cite{keyes2013multiphysics}. In many computational frameworks, solid or discrete regions are represented on regular grids or point sets that lack a well-defined interface~\cite{silling_pericode_2003,eshraghi2012implicit}, while continuum solvers rely on boundary-conforming geometries to evaluate surface fluxes~\cite{verzicco2023immersed,candler2019rate}. The mismatch between these representations introduces errors in geometric fidelity and flux transfer, particularly when the interacting solvers operate at different spatial and temporal resolutions~\cite{hong_toward_2024}. Existing approaches for interface treatment often constrain the surface description to the underlying discretization~\cite{hong_toward_2024,hong2025general} or limit the coupling formulation to a single class of physical phenomena, reducing their general applicability~\cite{mcquaid2025development}. To address these limitations, a framework is required that enables accurate interface reconstruction and robust data exchange between discrete solid and surface-based continuum domains.

Solid domains are often represented in scientific simulation by using a regular grid of rectangular elements. In the current work, the term voxel will be used to refer to these rectangular elements, whether they are two or three-dimensional. Usually these elements are square in 2D and cubic in 3D, but other aspect ratios can be used. Voxels are widely used as the building blocks when simulating larger geometries. In this article, we refer to geometries built using voxels as voxelized structures. In medical applications, voxelized structures are commonly used to accurately represent human anatomy~\cite{caon_voxel_2004, mcqueen_cardiac_2000, wang_3d_2021}. Radiation doses are calculated using Monte Carlo simulations that include voxelized approximations of human tissue~\cite{price_voxelization_2023, kramer_max_2004, cordeiro_dosimetry_2023}. Voxelized Monte Carlo has also been used in the study of photon-based medical imaging~\cite{periyasamy_montecarlo_2017, yan_photon_2020, leino_valo_2019}. In manufacturing, voxels can be used to model design geometries in additive manufacturing~\cite{ueng_voxel_2018, aremu_print_2017, tedia_manuf_2016} and to plan tool paths in subtractive manufacturing~\cite{lynn_machining_2017, zhang_voxel_2021, lynn_direct_2017, yu_model_2017}.

\begin{figure}[ht!]
\centering
    \includegraphics[width=0.95\textwidth, trim={0.1cm 0.1cm 0.1cm 0.1cm}, clip]{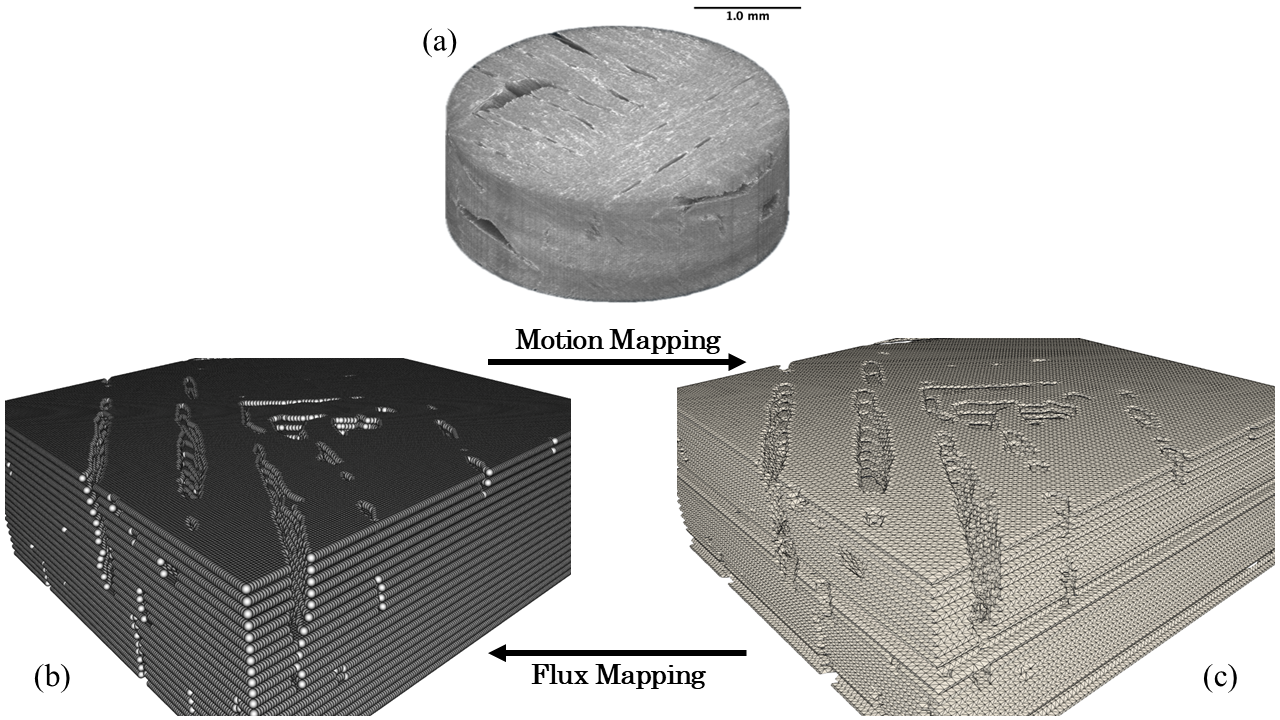}
\hfill
\caption{A voxelized carbon microstructure generated using XRCT scan data obtained at the University of Kentucky. A carbon composite sample was scanned and processed into the voxelized structure in (a). A zoomed-in portion of this voxelized sample material can be seen in (b). A boundary surface for the material portion in (b) is shown in (c).} 
\label{xrct}
\end{figure}

Engineering disciplines utilize a wealth of simulation techniques based on voxelized structures. For example, voxelized geometries are utilized in simulating solid structural response with techniques that include the lattice particle method (LPM)~\cite{chen3d} and peridynamics~\cite{silling_peridynamics_2000, silling_pericode_2003, silling_peristates_2007}. Both of these are meshfree structural simulation techniques in which the material points are represented by voxels or similar discrete elements. Voxel-based methods are also available to simulate heat transfer within systems like an inhabited building~\cite{goldstein_towards_2014, xu_point_2021, tyc_arch_2023}. The lattice Boltzmann method (LBM)~\cite{perumal_lbm_2015, aidun_lbm_2010, huang_lbm_2015} is a voxel-based simulation technique which is capable of capturing both fluid flow and heat transfer behaviors. Regions of the fluid are divided across a square lattice, and each voxel has a distribution function specified for its particles. The fluid lattice is simulated in LBM by alternately colliding particles within a voxel and advecting particles between voxels. The LBM technique has also been utilized in solid-liquid phase change simulations~\cite{li_pore_2018}. In the aerospace field, voxelized structures are generated to study thermal protection system (TPS) materials. Hypersonic vehicles require a robust method of temperature regulation to manage the high heat loads encountered at their leading edges. Carbon materials with porous microstructures are often used as sacrificial TPS material on hypersonic vehicles~\cite{mansour_ablative_2024, panerai_microtom_2017}. Image stacks generated from X-ray computed tomography (XRCT) can be used to provide accurate voxelized representations of carbon microstructures~\cite{mansour_ablative_2024}. These voxelized structures can then be incorporated into fluid simulations~\cite{chacon_rtv_2024, ramu_supervised_2023, poovathingal_permeable_2022} or into thermal simulations~\cite{liu_mct_2024}. Figure~\ref{xrct} illustrates an example of a voxelized structure generated from XRCT scans of a carbon composite sample. The entire voxelized sample is shown in Fig.~\ref{xrct}a, where both solid material and pores are visible in the structure. A zoomed-in voxelized portion of the sample can be seen in Fig.~\ref{xrct}b. 

One limitation of voxelized structures, whether used for ablation or any other application, is that they do not have a well-defined bounding interface. In this article, we refer to this bounding interface as the surface. When coupling traditionally voxel-based methods with other techniques in a multiphysics simulation, a true surface definition is required. For example, fluid flows are sensitive to even small changes in the geometry of immersed surfaces. Therefore, computational fluid dynamics~\cite{mark_immersed_2011, kong_sensitivity_2018} and direct simulation Monte Carlo (DSMC)~\cite{zhang_robust_2012} solvers generally need an accurate interface to calculate surface fluxes. Historically, marching cubes~\cite{lorensen_marching_1987} and its 2D analog marching squares~\cite{maple_geometric_2003} have been used to generate surfaces from voxels, but they are known to generate spurious surface features~\cite{hong_toward_2024}. In this work, the process of surface generation is referred to as motion mapping.

In addition to surface generation, if a simulation is coupled, then there must also be a way to accurately transfer data from the generated surface back to the underlying voxelized structure. For example, a fully coupled fluid-structure interaction (FSI) simulation would need to take surface forces applied by the flow and transfer them to the voxelized structure. These transferred forces act as a necessary boundary condition upon the structural simulation. This principle holds for other fully coupled multiphysics simulations as well. Thermal response simulations would require surface heat fluxes to be transferred so that conduction can occur through voxels. Simulating oxidation of carbon TPS materials would require surface reaction data to be transferred so that voxel masses can be ablated. The process of transferring surface data to voxels is referred to in this work as flux mapping. As is indicated in Fig.~\ref{xrct}, flux mapping is a method to transfer data from a boundary surface to a voxelized representation of the same geometry.

Two-way coupling of meshfree methods such as peridynamics and smoothed particle hydrodynamics (SPH) have been developed in the past~\cite{sun_fsi_2020, ren_peridynamicssph_2015}. Kim, Bhalla, and Griffith~\cite{kim_ipd_2023} coupled peridynamics with the incompressible Navier-Stokes equations using the immersed boundary method. Recently, Hong et al.~\cite{hong_toward_2024} presented a method which includes both motion and flux mapping procedures. They used a modified marching cubes procedure~\cite{lorensen_marching_1987} to generate surfaces for DSMC simulations. These surfaces were produced using voxelized structures that were defined on an underlying static grid. Surface fluxes representing material ablation were mapped from this surface to nearby voxels, allowing the surface to recede as the voxels ablated away. In this approach, the resolution of the voxelized structure is fixed to the resolution of the surface grid. This fixing of resolutions limits the applicability of the method to other types of multiphysics simulations. Surface elements should be sufficiently resolved to capture geometric features, but this resolution is not necessarily the same as that required to simulate the voxelized domain. Additionally, the flux mapping described in Ref.~\cite{hong_toward_2024} is only applicable to ablation problems. It is desirable to have a voxel-to-surface coupling method that can be used in a broad range of multiphysics simulation problems.

In this article, a new coupling method is described that relates a voxelized structure with a corresponding boundary surface. Novel motion and flux mapping procedures are developed that allow consistent coupling of partitioned systems, enabling complex multiphysical simulations. Early versions of this coupling method were used in FSI simulations described in Refs.~\cite{huff_fsi_2024, huff_fsi_2025}. In this method, the resolution of the voxelized structure and the surface grid are independent, and the flux mapping procedure can be used for mapping a wide variety of surface quantities to voxels. During flux mapping, conservation is ensured when scalar values on the boundary surface are transferred to the underlying voxelized structure. We refer to this two-way coupling method as the marching windows method.

\section{Marching Windows}
\label{sec_mwind}

\begin{figure}[ht!]
\centering
    \includegraphics[width=0.95\textwidth, trim={0.1cm 0.1cm 0.1cm 0.1cm}, clip]{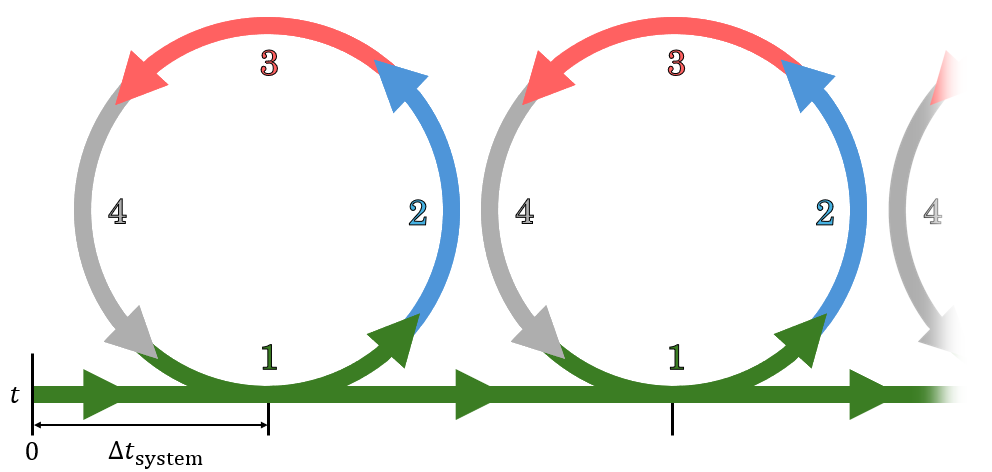}
\hfill
\caption{Coupling framework with marching windows. The framework consists of a repeating cycle of four stages, simulated with a time interval $\Delta t_\textnormal{system}$. (1) Generate surface from voxels using marching squares. (2) Simulate the fluid stage for a simulated time of $\Delta t_\textnormal{system}$. (3) Transfer fluxes from surface elements to voxels. (4) Simulate the solid stage for a simulated time of $\Delta t_\textnormal{system}$. In strongly coupled frameworks, cycles may be reiterated until converged. Once the cycle is complete, a new cycle begins with the generation of a surface from the new voxel structure.} 
\label{fsi_coupling_scheme}
\end{figure}

Marching windows is a pair of modules designed to couple two disparate simulation types, one type that uses a voxelized structure and the other that requires a bounding surface. The motion mapping module is a set of procedures which takes a voxelized structure and generates a corresponding boundary surface. The flux mapping module relates each surface element in the boundary to nearby voxels to facilitate conversion of surface-based data to voxel-based data. Figure~\ref{fsi_coupling_scheme} illustrates a schematic of a generic framework for coupled multiphysics simulations that uses marching windows. The illustrated framework includes both a solid domain consisting of voxels and a fluid domain that requires a bounding surface. Even though the ideas of marching windows are not strictly limited to fluid-solid multiphysics frameworks, the terminology used in this article will reflect a fluid-solid coupling framework for clarity. A full system cycle in the coupled framework consists of four distinct stages which simulate the coupled system across the duration of a system timestep ($\Delta t_\textnormal{system}$). Two of the stages are time-marching stages for the fluid domain and solid domain. The other two stages consist of data transfer between the two domains using the marching windows modules.

The system cycle begins with a voxelized solid structure. This initial structure could be a carbon microstructure from a TPS, a block of metal through which heat conducts, or any other system that is representable with a set of voxels in space. Stage one in Fig.~\ref{fsi_coupling_scheme} is motion mapping, beginning at time $t = 0$. Motion mapping generates an outer boundary surface for the structure, and that boundary is immersed into the fluid domain. The details of the motion mapping procedures are explained in Sec.~\ref{sec_motionmap}. Stage two is fluid time-marching. Using the boundary generated in stage one, the fluid is simulated for the duration of the system timestep. Stage three consists of flux mapping. Flux mapping converts the relevant surface data generated in stage two to voxel quantities to act as boundary conditions in the solid simulation. The physical meaning of this surface data will vary depending on the type of simulation. Possible candidates for surface data include aerodynamic forces, heat transfer, and mass fluxes from ablation. The flux mapping module is detailed in Sec.~\ref{sec_fluxmap}. Stage four is solid time-marching. The solid is simulated with the boundary conditions calculated in stage three for the duration of a system timestep.

Once the solid has been simulated in stage four, the next procedure depends on whether the framework is strongly or loosely coupled, as indicated by the branching paths in Fig.~\ref{fsi_coupling_scheme} after stage four. In a loosely coupled framework, the path towards the next cycle is taken and the next system timestep begins immediately from stage one. In this new cycle, the motion mapping module regenerates the boundary surface using the new voxelized structure. Changes in voxelized geometry can result from mechanical deformation, ablation, or other causes relevant to the solid simulation. In any case, the boundary from the previous system cycle is updated with the new geometry of the structure. The cycle then continues as before with stages two, three, and four, and subsequent system timesteps proceed similarly.

In a strongly coupled framework, the same cycle may be repeated to achieve convergence before the next system cycle begins. In the model of Fig.~\ref{fsi_coupling_scheme}, the coupled simulation begins at time $t = 0$, and the first cycle begins with stage one. A boundary surface is generated in stage one, and the rest of the cycle proceeds as described above. Once the cycle ends, the cycle can be re-simulated using the new voxel configuration resulting from stage four. This repetition corresponds to taking the path back into the cycle in Fig.~\ref{fsi_coupling_scheme} after stage four. Repetition continues until the results have converged to stable values. Once convergence is achieved, the path leading to the next cycle in Fig.~\ref{fsi_coupling_scheme} can be taken, and the next system timestep can be simulated. Each subsequent system timestep is similarly repeated until converged.

The marching windows modules allow two disparate simulation domains to share data consistently with each other. Motion mapping provides a surface to the fluid domain, and that surface acts as an interface that can transfer fluid data to the voxelized solid.


\FloatBarrier

\section{Motion Mapping}
\label{sec_motionmap}

\subsection{Motion Mapping Method}

The motion mapping module of the marching windows method takes a voxelized structure as input and outputs a bounding surface corresponding to that structure. The motion mapping process is composed of four steps. The first step is generating a Cartesian grid, the marching windows grid, that overlays on the voxelized structure. Each corner, or node, in the marching windows grid contains a scalar that represents a volume fill fraction in the vicinity of the node. The volume fill fractions are calculated in steps two and three of the motion mapping process. Step two involves applying a weighting function to each voxel depending on its position relative to the edge of the voxelized structure. The weight of each voxel is applied to its calculated volume (area with unit depth). The third step is using these weighted voxel volumes to calculate the volume fill fraction at each node in the marching windows grid. The fourth and last step is the application of marching squares~\cite{maple_geometric_2003} on the scalar field of volume fill fractions. Marching squares estimates with linear interpolation the isosurface of 50\% fill fraction in the marching windows grid, and this isosurface is the estimated bounding surface of the voxelized structure. Each of the four steps of motion mapping is detailed in order below. The current work focuses on 2D structures, but motion mapping can be translated into a 3D version as well. Each description of the steps below includes details on analogous 3D procedures which can be used in a 3D motion mapping module.

\subsubsection{Marching Windows Grid}

\begin{figure}[ht!]
\centering
\begin{subfigure}{0.32\textwidth}
    \centering
    \includegraphics[width=0.99\textwidth]{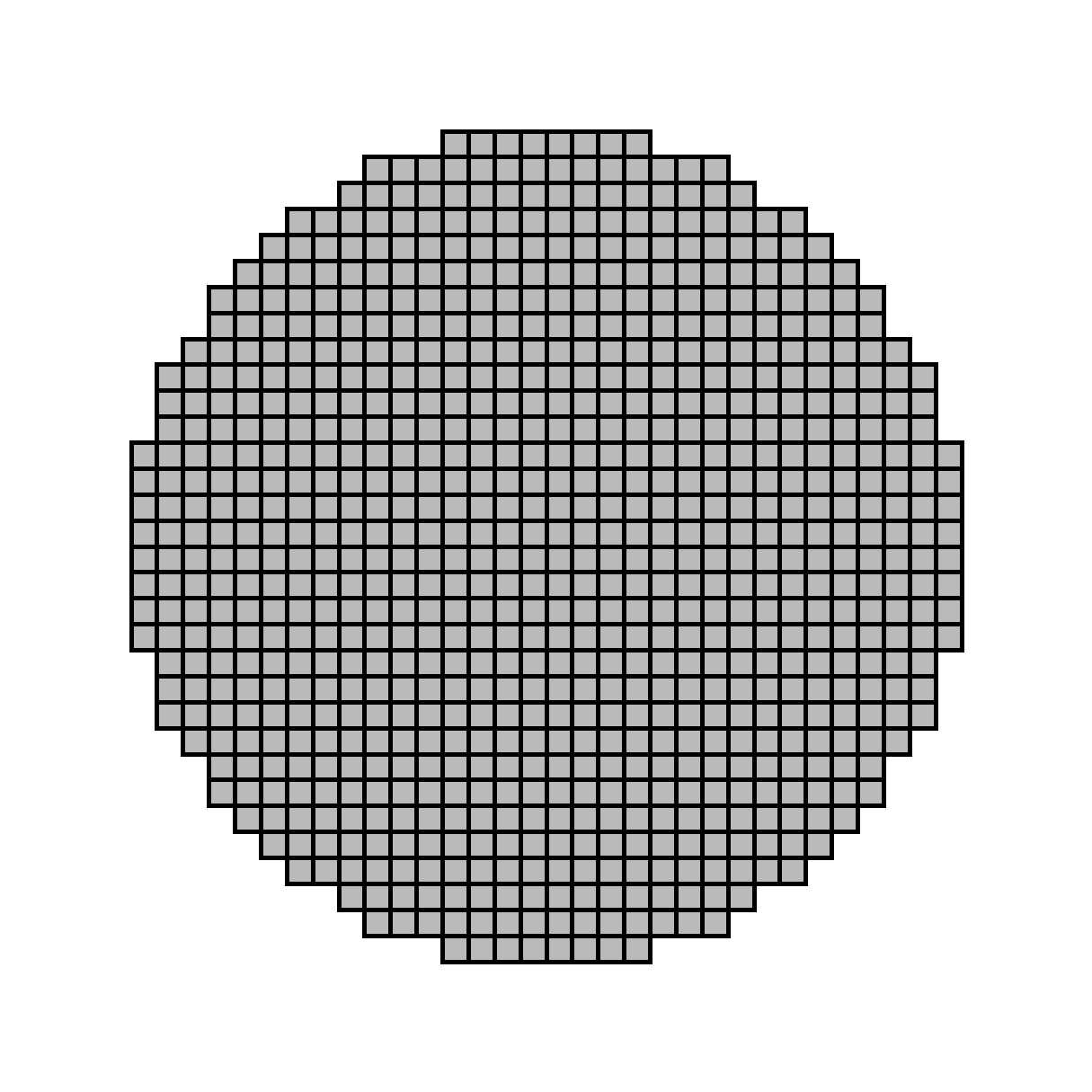}
    \caption{Voxelized circle.}\label{mwgrid_vox}
\end{subfigure}
\hfill
\begin{subfigure}{0.32\textwidth}
    \centering
    \includegraphics[width=0.99\textwidth]{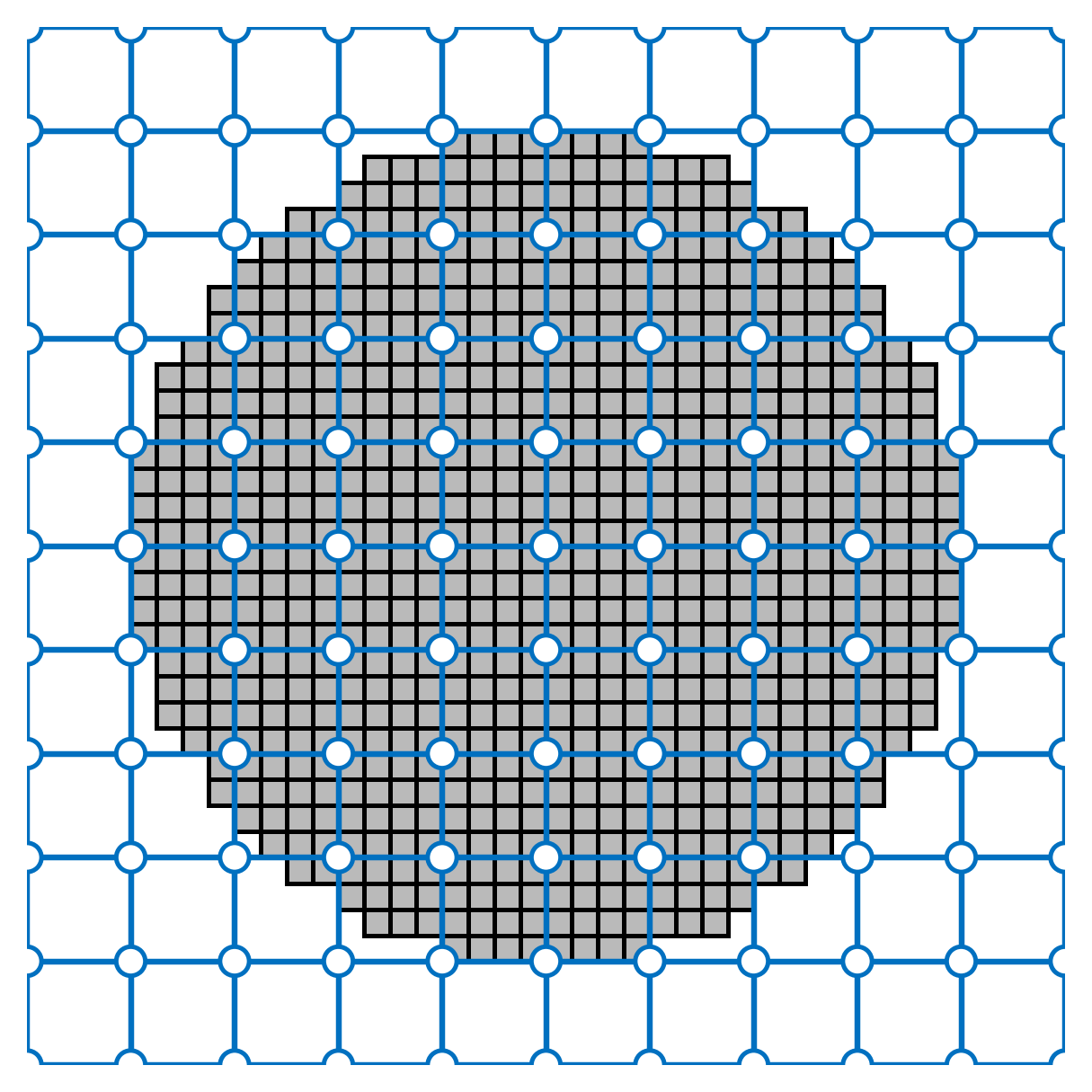}
    \caption{Coarse grid.}\label{mwgrid_cell}
\end{subfigure}
\hfill
\begin{subfigure}{0.32\textwidth}
    \centering
    \includegraphics[width=0.99\textwidth]{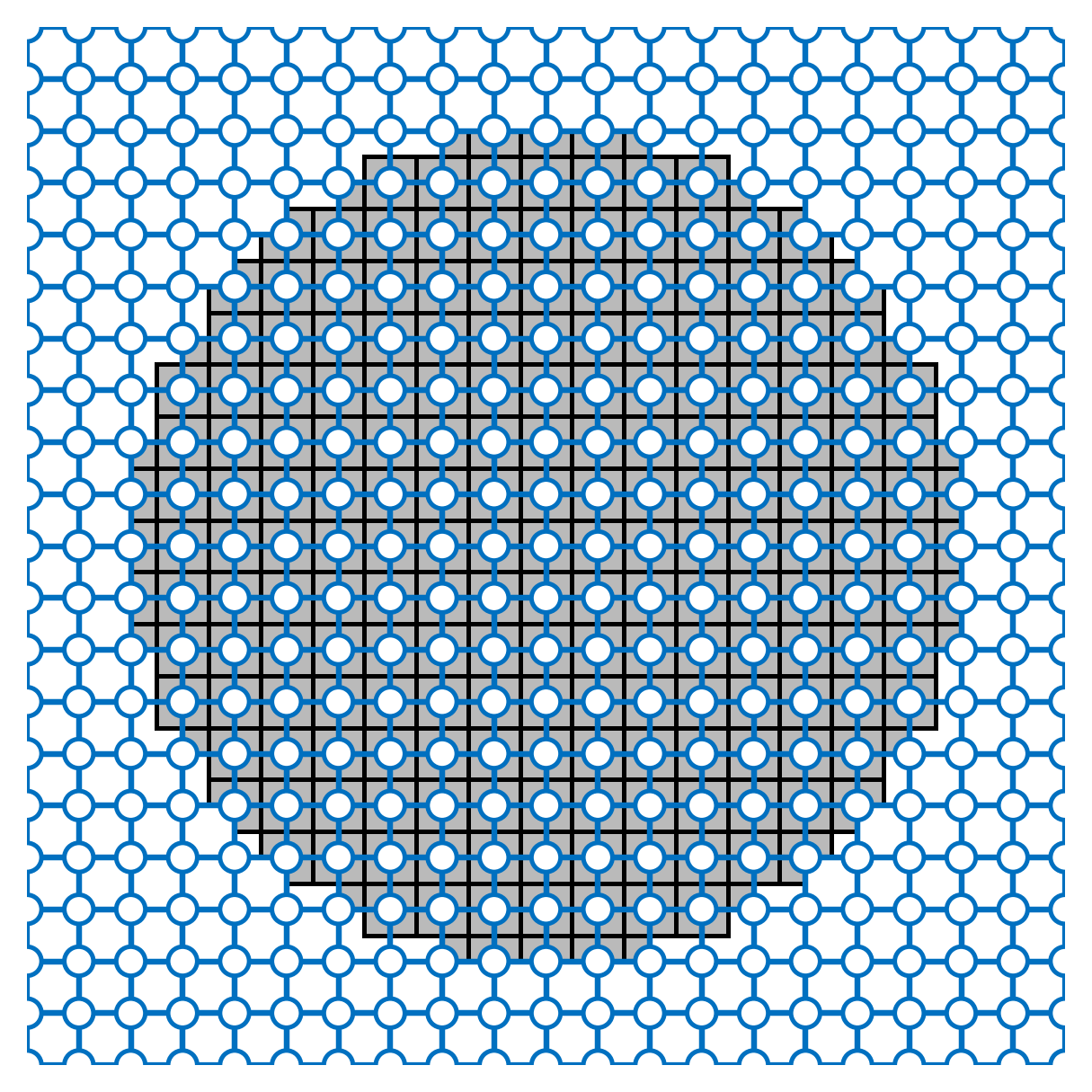}
    \caption{Refined grid.}\label{mwgrid_refine}
\end{subfigure}
\hfill
\caption{Overlaying of marching windows grid over a voxelized circle. The initial voxelized structure (a) is covered with a coarse marching windows grid (b) and a more refined grid (c). Grid cells are outlined in blue, and grid nodes are notated as white circles.} 
\label{mwgrid}
\end{figure}

The first step in the motion mapping procedure is to overlay the marching windows grid onto a pre-existing voxelized structure. The marching windows grid is a uniform, square Cartesian grid. Two examples of these grids are illustrated in Fig.~\ref{mwgrid}. Figure~\ref{mwgrid_vox} illustrates a circular structure composed of gray voxels which will serve as an example geometry for the motion mapping procedure. Two examples of marching windows grids are given in Fig.~\ref{mwgrid_cell} and Fig.~\ref{mwgrid_refine}. Each square formed by four blue edges is a cell in the grid. The corners of the grid, denoted with the white circles, are the nodes that will contain the volume fraction scalars for marching squares later in the motion mapping procedure.

In order to define the marching windows grid, the required parameters are bounding box limits and the side length of the cells. The limits are the minima and maxima of the grid node coordinates. These limits should be chosen such that voxels are not near the boundaries of the grid. The details of selecting grid limits are discussed in Sec.~\ref{sssec_gvoxels}. The grid of Fig.~\ref{mwgrid_cell} uses a cell size equal to four voxel widths, while the more refined grid of Figure~\ref{mwgrid_refine} uses a cell size of two voxel widths. In a 3D version of marching windows, extending this grid overlaying step would merely require using cubic cells and using 3D coordinates for the bounding box limits.

In order to keep the discussion of marching windows nondimensional, voxel size ($L_v$) and cell size ($L_c$) will be described in the rest of this work relative to the size of the geometry under consideration. All geometries under consideration are polygons or circles, so the circumradius ($R_\text{circ}$) of each geometry is used as the characteristic length. Voxel sizes are defined with a voxel ratio ($v_r$),
\begin{equation}
    v_r = \frac{R_\text{circ}}{L_v}
\end{equation}
and cell sizes are defined with a grid ratio ($g_r$),
\begin{equation}
    g_r = \frac{R_\text{circ}}{L_c}
\end{equation}
For example, if $v_r = g_r$, voxel size is equal to cell size and one voxel can fit in each cell. If $v_r = 2g_r$, then four voxels can fit in each cell.

In the current work, a few limitations are applied to the grids used. First, each grid is aligned with the corresponding voxelized structure. The nodes of the grid are coincident with corners of voxels as in Fig.~\ref{mwgrid_cell} and Fig.~\ref{mwgrid_refine}. Second, the cell size is set as a whole number multiple of voxel size. These grid limitations are not strictly necessary for the motion mapping procedures described in this section but are used for the error analysis discussed in Sec.~\ref{ssec_error}. Additionally, only cell sizes greater than or equal to voxel sizes are considered ($g_r \leq v_r$). As voxel size approaches and exceeds cell size ($g_r > v_r$), surface roughness resulting from the sharp, rectangular geometry of the voxels becomes more evident. An example of this roughness is discussed in Sec.~\ref{ssec_error}.

\subsubsection{Ghost Voxels and Weighting}
\label{sssec_gvoxels}

\begin{figure}[ht!]
\centering
\begin{subfigure}{0.533\textwidth}
    \centering
    \includegraphics[width=0.99\textwidth]{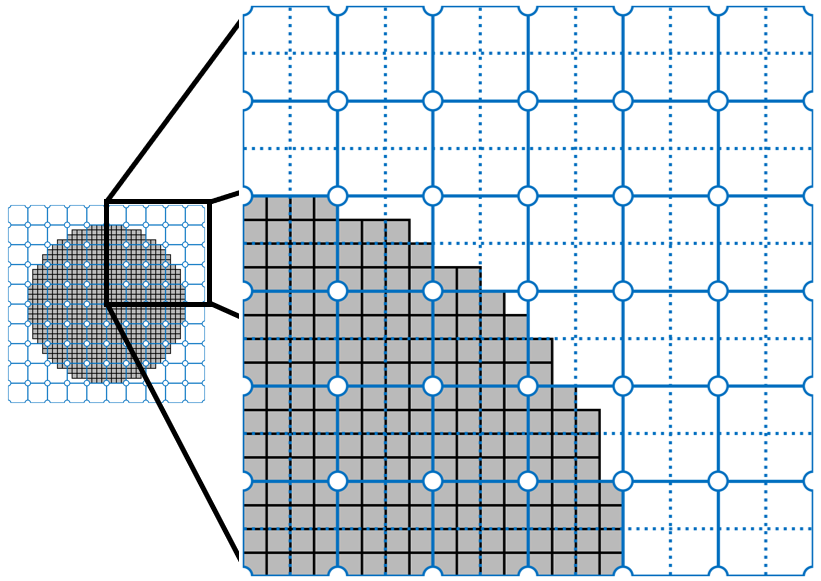}
    \caption{Top-right quadrant of circle.}\label{mmweight_regions}
\end{subfigure}
\begin{subfigure}{0.455\textwidth}
    \centering
    \includegraphics[width=0.99\textwidth]{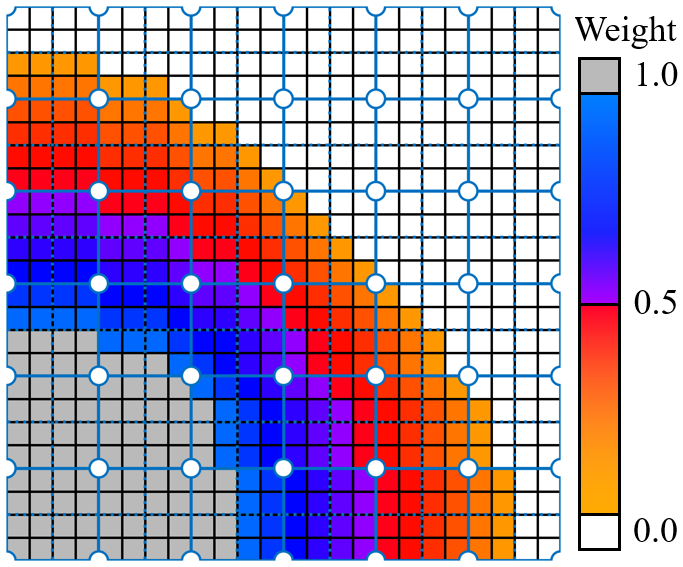}
    \caption{Voxel weights of circle quadrant.}\label{mmweight_color}
\end{subfigure}
\caption{Nodal volume division and voxel weighting.} 
\label{mmweight}
\end{figure}

The second step in motion mapping is the weighting of voxel volumes before distributing them to grid nodes. The voxel weighting is illustrated in Fig.~\ref{mmweight}. Figure~\ref{mmweight_regions} displays the top-right quadrant of the grid from Fig.~\ref{mwgrid_cell}. The blue dotted lines indicate midlines between two nodes in the horizontal or vertical direction. Each node owns the volume enclosed by the square of four dotted edges surrounding it, i.e. each node owns a nodal volume of $L_c^2$ centered on itself. The volume fill fraction of each node is calculated by the fraction of its nodal volume that is filled by voxel volumes.

The weights that are applied to voxels before the nodal volume fractions are calculated are illustrated in Fig.~\ref{mmweight_color}. A weight of 1.0 (gray) means that the entire volume of the voxel is considered, while a weight of 0.0 (white) means none of the voxel volume is considered in the nodal fill fraction. Weights in between these two extremes indicate that a fraction of the voxel volume is included in the filled nodal volume. For example, the nodes at or near the center of the circle that are completely filled with gray voxels are 100\% filled. The nodes near the edge voxels (bright purple) are between 0\% and 100\% filled.
Voxel volumes are weighted to address an artifact of linearly interpolating volume fill fractions to generate surfaces. This surface artifact is detailed in Sec.~\ref{sssec_msquares}.

The weighting function ($w_v$) used for each voxel is a function of its depth in the structure. It is expressed as
\begin{equation}
    w_v = \frac{1}{2} + \frac{(\frac{1}{2} + d_v)L_v}{3L_c}
    \label{eq_vox_weight}
\end{equation}
where $d_v$ is the integer depth of the voxel. Edge voxels have depth $d_v = 0$ and are defined as any voxels which have a face exposed to empty space. The depth of interior voxels is determined by the number of steps on the shortest path from their original positions to an edge voxel, where valid steps are from the current position to a position that shares a face with it. Calculating depth is analogous to counting how many squares a chess rook would need to traverse from the current position to the nearest edge voxel. The method of depth counting is illustrated at the bottom of Fig.~\ref{mmweight_color}. The bottom, bright purple voxel is the edge at depth $d_v = 0$, located in the same position as in Fig.~\ref{mmweight_regions}. The voxel just interior to it (to the left) is at depth $d_v = 1$. The next voxel to the left is at depth $d_v = 2$, and so on.

The function in Eq.~\ref{eq_vox_weight} was chosen such that weight would be set at 50\% at the outer edge of a flat wall of voxels. For reasons discussed in Sec.~\ref{sssec_msquares}, this weighting allows generated boundary surfaces to exactly coincide with the outer edge of the voxel wall. To ensure that boundary surfaces and voxel wall edges are coincident, empty spaces near the edge voxels must be treated as low-weight voxels as well. Therefore, ghost voxels are generated from the empty space around the voxelized structure. These ghost voxels only exist for the purpose of volume weighting in motion mapping, and their weights are set by Eq.~\ref{eq_vox_weight} as for regular voxels. The depth counting scheme also works in the same manner as for regular voxels. True voxels at the edge still have depth $d_v = 0$. The ghost voxels in the first layer outward from the edge have depth $d_v = -1$, those of the second layer have depth $d_v = -2$, and so on. Note, for both true and ghost voxels, there is no need for voxel weights above one or below zero. Therefore, the weights of any voxels that would exceed those bounds are clipped,
\begin{equation}
    0 \leq w_v \leq 1
\end{equation}
In Fig.~\ref{mmweight_color}, the transition from deep interior voxels, through weighted voxels, through ghost voxels, and out to empty space is illustrated. The deep interior voxels have a weight capped at one. As one moves from the deep interior towards the edge of the solid, weights begin to decrease from one. This initial decrease is illustrated by the gradient of cool colors in Fig.~\ref{mmweight_color}. The first layer of ghost voxels is visible as the first red layer outside the purple edge voxels. The ghost voxel weights decrease with their distance from the edge. This decrease is visible in the gradient of warm colors. At the end of the warm gradient where the voxel weight goes to zero, the empty voxel spaces are marked in white. In order for the marching windows grid nodes to capture all true and ghost voxel volumes, the bounding box limits of the grid should be set sufficiently far from any true voxels to prevent ghost voxels from penetrating outside of the box. Based on Eq.~\ref{eq_vox_weight}, the depth at which ghost voxel weight reaches zero ($d_{v,\text{min}}$) is
\begin{equation}
    d_{v,\text{min}} = -\frac{3}{2}\frac{L_c}{L_v} - \frac{1}{2}
\end{equation}
where $d_{v,\text{min}}$ is not necessarily an integer. This depth provides a conservative estimate for the distance that non-zero-weighted ghost voxels can reach beyond the edge voxels. Since $d_{v,\text{min}}$ is the negative of the number of voxels that can reach outwards in a straight line from the true voxel edge, the farthest distance which a ghost voxel can reach ($l_\text{max}$) is
\begin{equation}
    l_\text{max} = \frac{3}{2}L_c + \frac{1}{2}L_v
\end{equation}
Therefore, $l_\text{max}$ can be used to set the minimum size of the marching windows grid based on the bounding box of the voxelized structure. To keep ghost voxels inside the marching windows grid, the bounding box of the structure should be completely enveloped by the bounding box of the grid with a minimum buffer between the two boxes of $l_\text{max}$ at all points.

The voxel weighting step in the motion mapping procedure can be performed in a 3D framework using the same logic for depth and weight. For edge voxels, instead of having four possible exposed faces, there would be six. The depth calculation would be done in the same way as in 2D by counting the number of steps each voxel needs to traverse to reach the nearest edge voxel through faces. Ghost voxels would be produced using the same depth traversal procedure as true voxels.

\subsubsection{Voxel Volume Distribution}

\begin{figure}[ht!]
\centering
\begin{subfigure}{0.462\textwidth}
    \centering
    \includegraphics[width=0.99\textwidth]{mm_procedure/mm_2b.png}
    \caption{Weighted voxels.}\label{mmvoldist_vox}
\end{subfigure}
\begin{subfigure}{0.465\textwidth}
    \centering
    \includegraphics[width=0.99\textwidth]{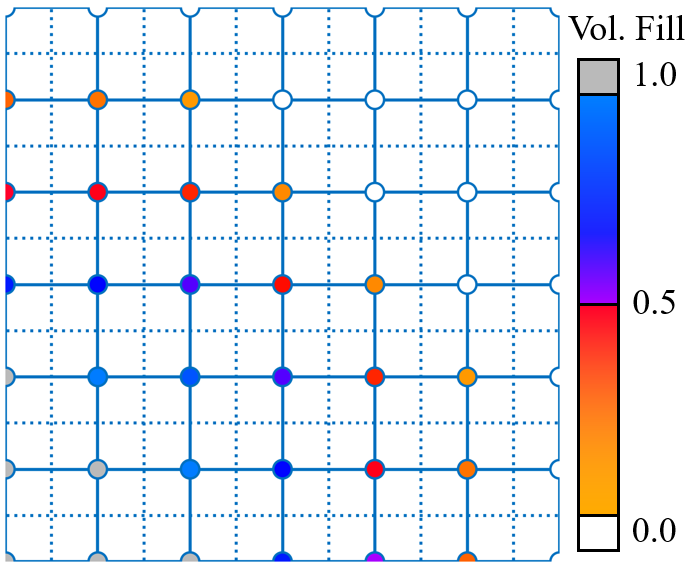}
    \caption{Nodal volume fill fractions.}\label{mmvoldist_nodes}
\end{subfigure}
\caption{Voxel volume distribution to marching windows grid nodes.} 
\label{mmvoldist}
\end{figure}

The third step in motion mapping is to distribute each of the voxel volumes to the corresponding marching windows grid node. This step is illustrated in Fig.~\ref{mmvoldist}. Figure~\ref{mmvoldist_vox} is Fig.~\ref{mmweight_color} reprinted for context. The volumes of the weighted voxels represented in Fig.~\ref{mmvoldist_vox} are divided among the nodes according to position. Since all voxels in Fig.~\ref{mmvoldist_vox} are each uniquely contained in one nodal volume, then the total volume fill fraction ($V_{\text{fill},\nu}/V_\text{node}$) of each node ($\nu$) from all $n_{\text{vox},\nu}$ owned voxel volumes is
\begin{equation}
    \frac{V_{\text{fill},\nu}}{V_{\text{node}}} = \frac{\sum_i^{n_{\text{vox},\nu}} w_{v,i} L_v^2}{L_c^2}
\end{equation}
where $w_{v,i}$ is the weight of voxel $i$. The results of the volume distribution to nodes are shown in Fig.~\ref{mmvoldist_nodes}. The color scheme for nodal volume fill fraction is the same as that for voxel weighting.
In a 3D version of marching windows, the distribution process would be the same. The only difference would be that the nodal volumes and voxels are cubes rather than squares with unit depth.

In the current analysis, we limit ourselves to voxels that are wholly contained in one nodal volume. In the case of Fig.~\ref{mmvoldist}, the entirety of each voxel volume falls within one nodal volume, i.e. there are no voxels which cross a blue, dotted line in Fig.~\ref{mmvoldist_vox}. In cases where voxels may cross nodal volume borders, the voxel volumes are divided between the nodes according to the fraction of the voxel residing in each nodal volume. For example, take a voxel with weight 0.7 that is positioned on a border between two nodal volumes with 40\% of the voxel on one side and 60\% on the other. In that case, 40\% of the voxel volume would go to the first node, and 60\% would go to the second node. Both of the contributions to the nodal volumes would be weighted with the original voxel weight of 0.7.

\subsubsection{Marching Squares}
\label{sssec_msquares}

\begin{figure}[ht!]
\centering
\begin{subfigure}{0.39\textwidth}
    \includegraphics[width=0.99\textwidth]{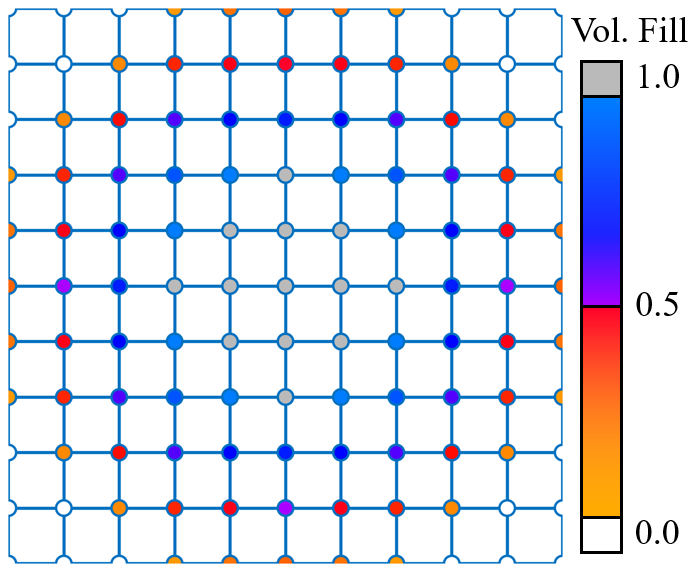}
    \caption{Nodal fill fractions.}\label{msquare_node}
\end{subfigure}
\begin{subfigure}{0.39\textwidth}
    \includegraphics[width=0.99\textwidth]{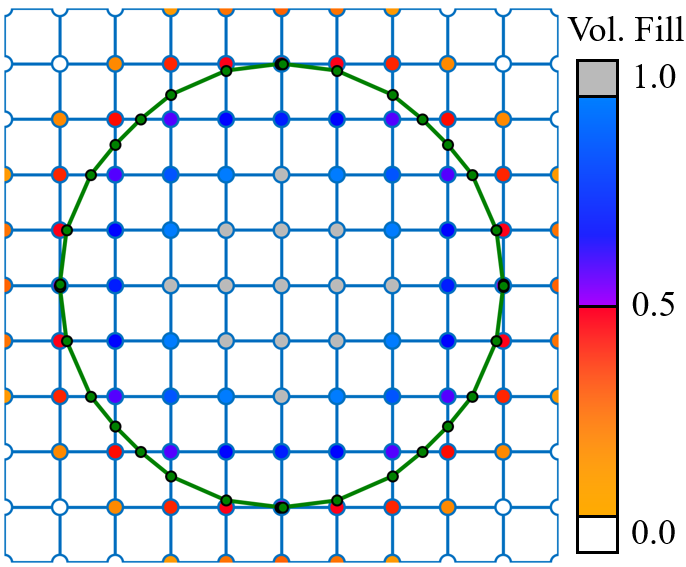}
    \caption{Nodes with surface.}\label{msquare_nodesurf}
\end{subfigure}
\begin{flushleft}
\hspace{0.095\textwidth}
\begin{subfigure}[b]{0.39\textwidth}
    \includegraphics[width=0.99\textwidth]{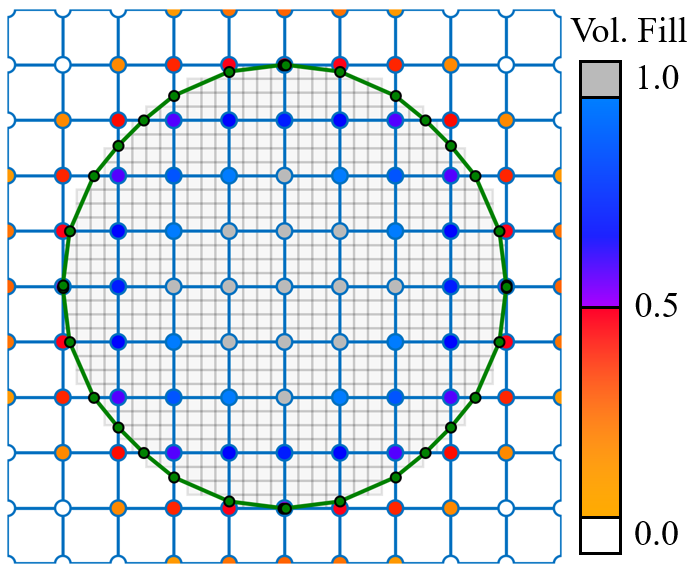}
    \caption{Nodes and surface with voxels.}\label{msquare_vox}
\end{subfigure}
\begin{subfigure}[b]{0.32\textwidth}
    \includegraphics[width=0.99\textwidth]{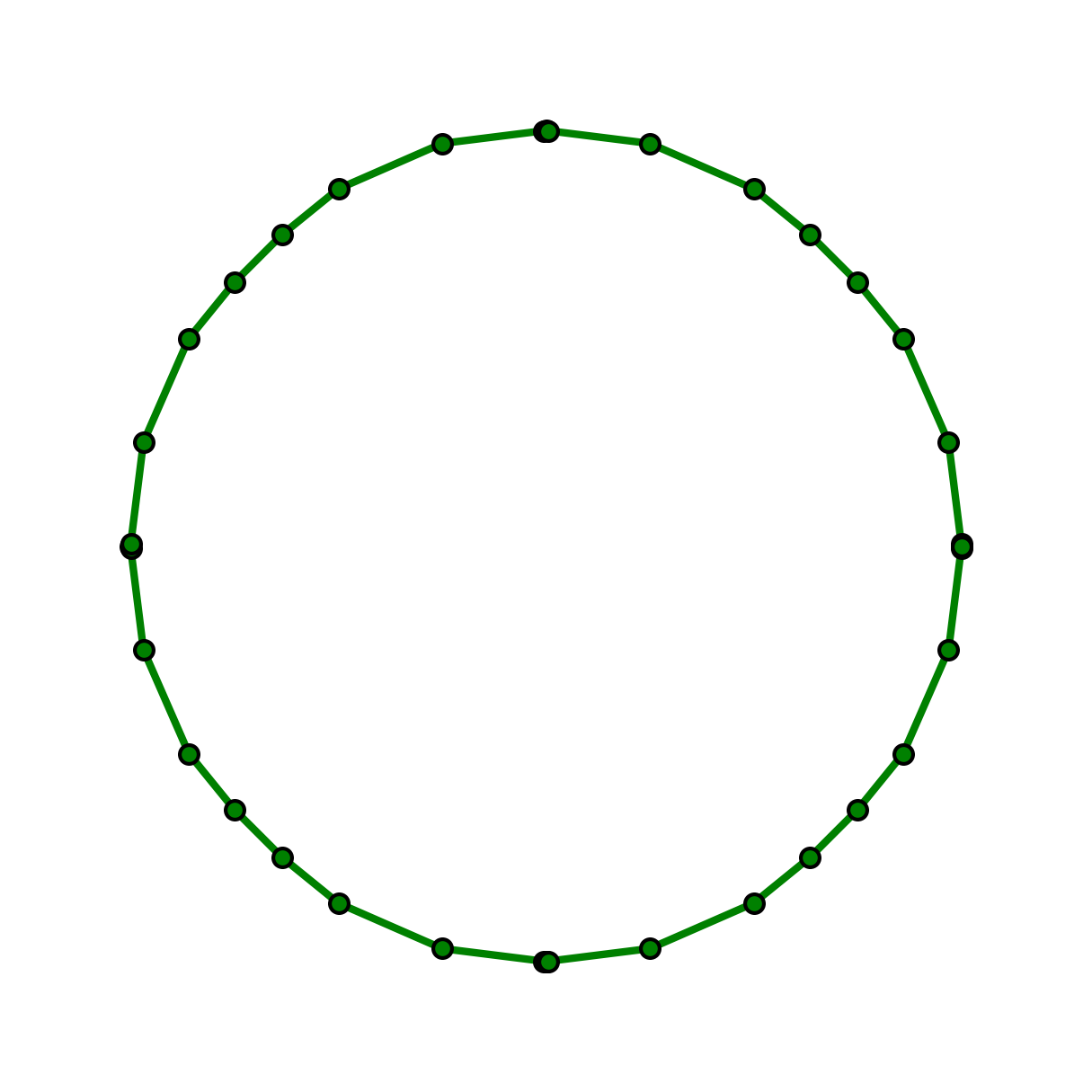}
    \caption{Marching squares surface.}\label{msquare_surf}
\end{subfigure}
\caption{Application of marching squares to a grid of nodal volume fill fractions.} 
\label{msquare}
\end{flushleft}
\end{figure}

The fourth and last step in motion mapping is the application of marching squares to the grid of volume fill fractions. This step produces a connected set of line segments that estimates the contour line of 50\% volume fill fraction in the grid. The original set of 16 cell topologies in marching squares~\cite{maple_geometric_2003} is used in this work. An exception is the saddle point topologies (indices 5 and 10), which are reversed to maintain connections between what may be otherwise separate geometries. Figure~\ref{msquare} illustrates the generation of a surface with marching squares.

Figure~\ref{msquare_node} is the full marching windows grid, color-coded by the nodal volume fill fractions calculated in the third step. Marching squares iterates through each cell in the marching windows grid and estimates the line (or lines) that represents the 50\% volume fill isosurface. First, a basic topology is chosen based on which nodes at the corners of the cell have a volume fill fraction greater than 50\% and which have a fill fraction less than 50\%. A line is drawn from one edge of the cell to another edge such that the less-than, or outside, nodes are separated from the greater-than, or inside, nodes by that line. The separation of outside and inside nodes is illustrated in Fig.~\ref{msquare_nodesurf}. In each cell that has a surface element (green), the element separates the outside nodes (orange and red) from the inside nodes (purple and blue). After the basic topology is applied, the surface element endpoints are adjusted by linear interpolation in order to more closely approximate the contour line at 50\% volume fill. For example, if a surface endpoint is placed on an edge whose nodes have 0\% and 75\% fill fractions, then the surface endpoint will be placed two-thirds of the way from the 0\% node to the 75\% node. The surface elements at the top, bottom, left, and right extrema of the surface in Fig.~\ref{msquare_nodesurf} lie very close to or directly on the node because that node is nearly or exactly 50\% filled. The nodes and surface from Fig.~\ref{msquare_nodesurf} are shown in Fig.~\ref{msquare_vox} with the original voxelized structure. The space enclosed by the surface captures the general shape of the voxelized structure, and the surface has a relatively smooth curvature. The surface is shown in Fig.~\ref{msquare_surf} by itself for clarity. A 3D version of marching windows could use the analogous marching cubes~\cite{lorensen_marching_1987} procedure in place of marching squares. Marching cubes generates triangular surface elements from cubic cells with volume fill nodes at the corners.

\begin{figure}[ht!]
\centering
    \includegraphics[width=0.95\textwidth, clip]{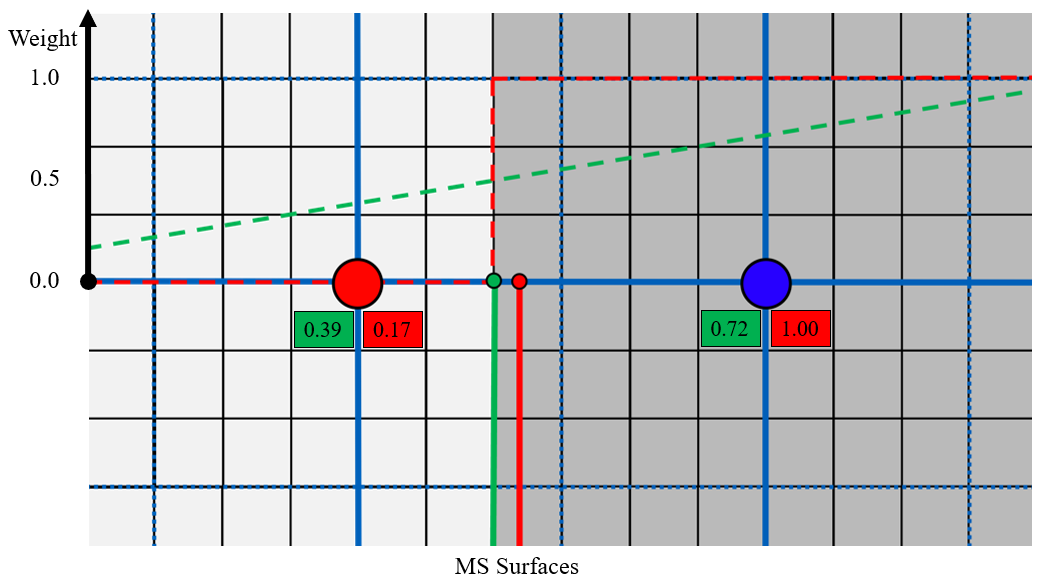}
\hfill
\caption{A linear interpolation artifact in unweighted marching windows. Two nodes are indicated, along with their weighted (green) and unweighted (red) volume fill fractions. The dashed green and red lines represent the voxel weighting functions, and the solid lines indicate the resulting surface elements.} 
\label{voxel_weight}
\end{figure}

The linear interpolation of marching squares is the reason that voxel weighting was used in step two, as there are cases where unweighted voxels can introduce surface imperfections. Figure~\ref{voxel_weight} compares a weighted method and an unweighted method of distributing voxel volumes to nodes. A wall of voxels extends up, down, and right, indicated in gray. The wall faces left towards the off-white open space. The weighting functions being compared are indicated by the dashed lines in the top half of Fig.~\ref{voxel_weight}. The dashed red line represents a binary weighting of zero for empty space and one for every voxel, and the dashed green line represents the weighting function of Eq.~\ref{eq_vox_weight}. When the binary weighting is used in this wall example, the resulting nodal fill fractions (17\% and 100\%) result in a surface element that is slightly offset from the edge of the voxelized structure. This offset surface element is indicated by the solid red line in the bottom half of Fig.~\ref{voxel_weight}. Ideally, the surface element in this flat wall geometry should be coincident with the actual edge of the voxelized structure. When using the weighting function of Eq.~\ref{eq_vox_weight}, the discrepancy between the generated surface element and the wall edge disappears. The nodal fill fractions resulting from the weighted voxels (39\% and 72\%) place the linearly interpolated 50\% contour line directly on the wall edge.

\FloatBarrier

\subsection{Error Analysis}
\label{ssec_error}

\begin{figure}[ht!]
\centering
\begin{subfigure}{0.32\textwidth}
    \centering
    \includegraphics[width=0.99\textwidth]{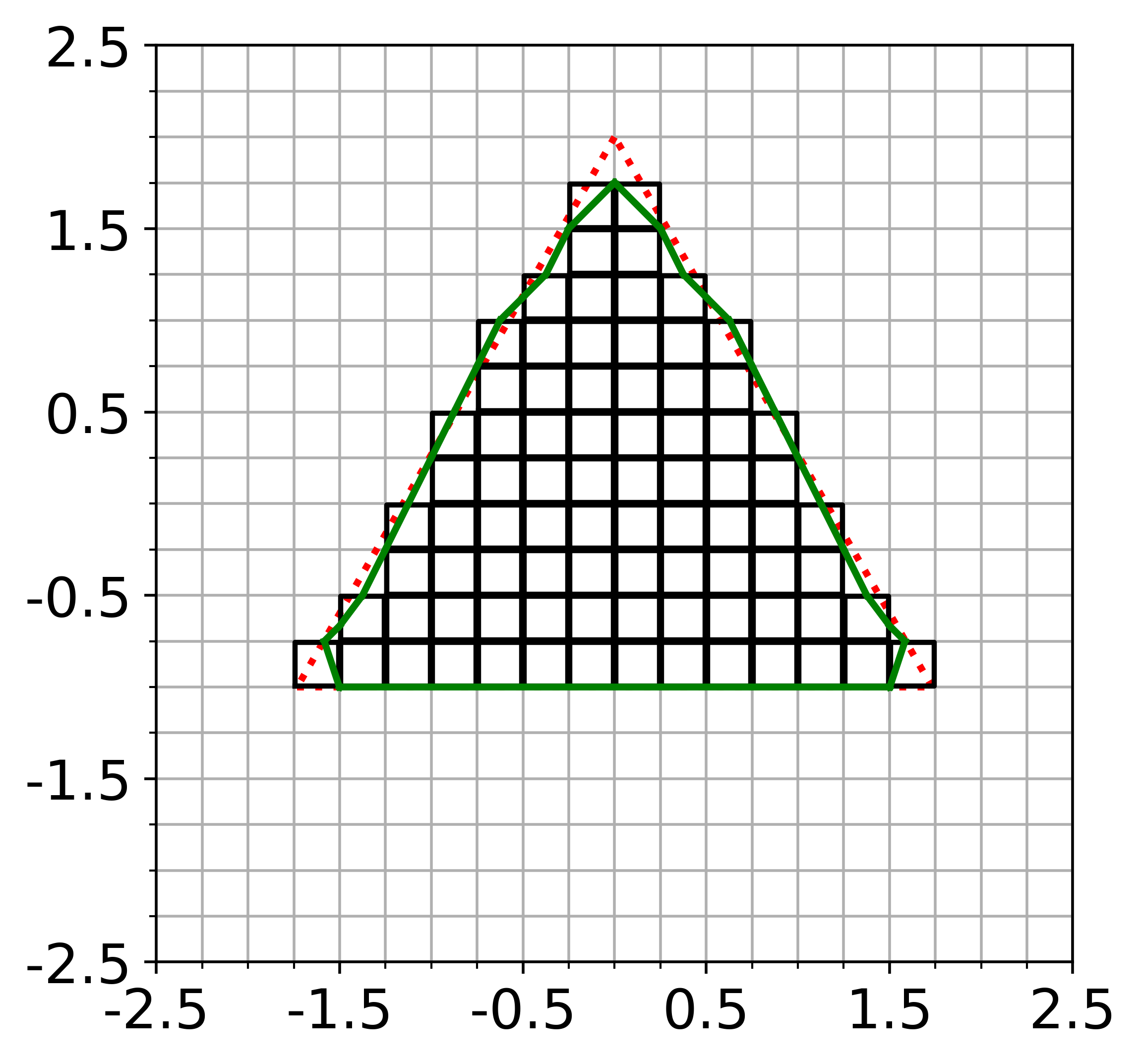}
    \caption{Triangle.}\label{mm_triangle}
\end{subfigure}
\hfill
\begin{subfigure}{0.32\textwidth}
    \centering
    \includegraphics[width=0.99\textwidth]{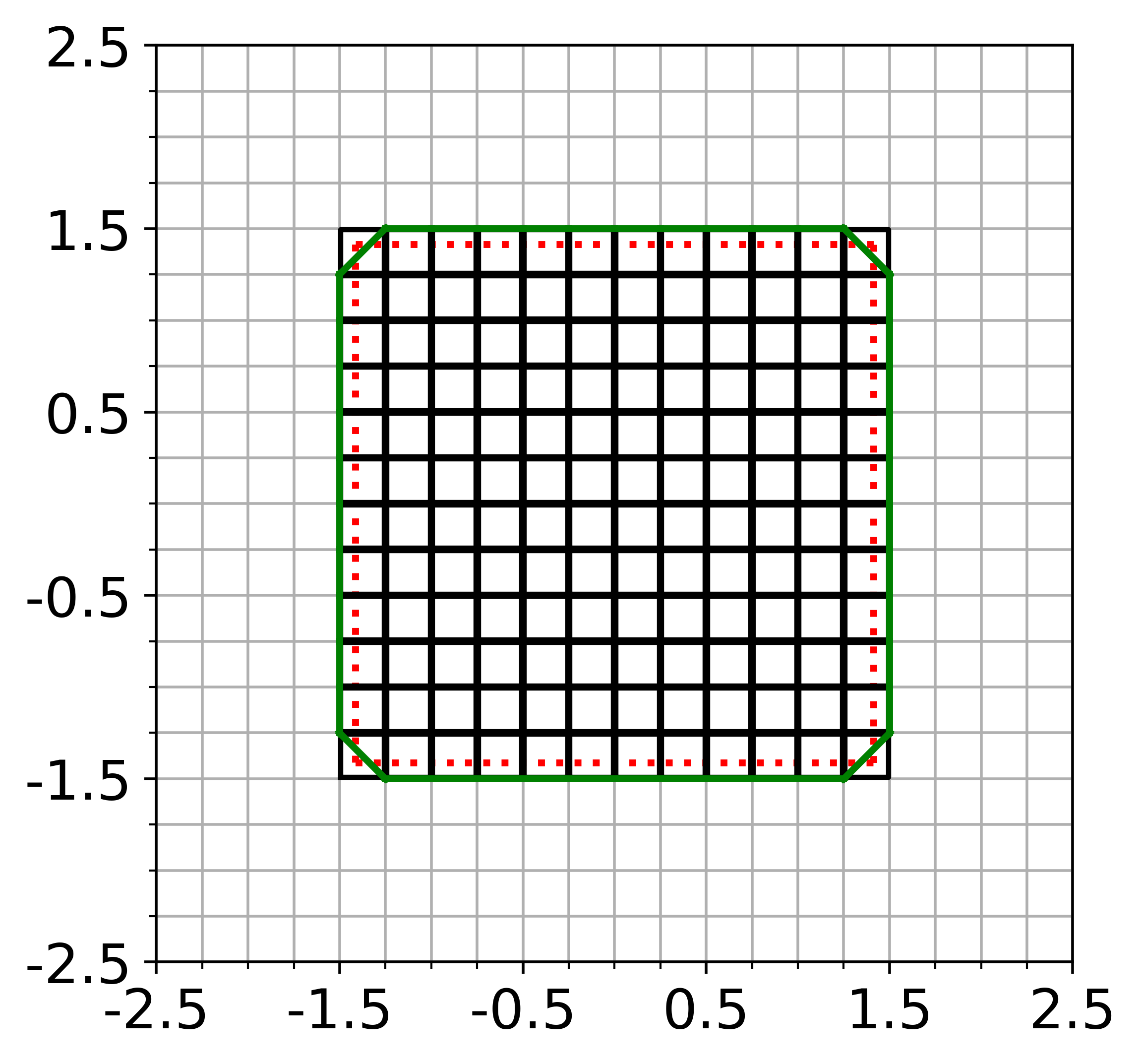}
    \caption{Square.}\label{mm_square}
\end{subfigure}
\hfill
\begin{subfigure}{0.32\textwidth}
    \centering
    \includegraphics[width=0.99\textwidth]{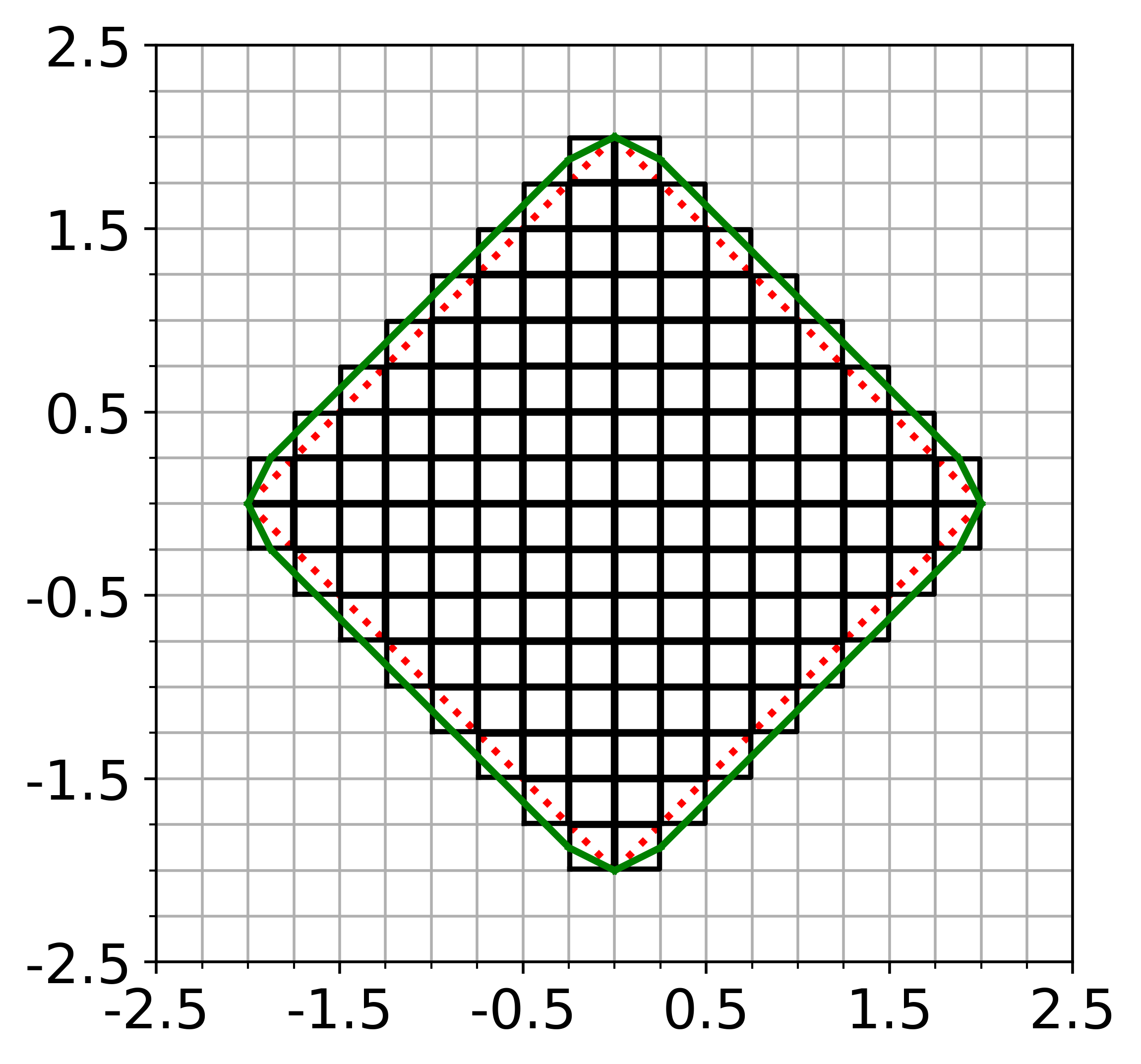}
    \caption{Diamond.}\label{mm_diamond}
\end{subfigure}
\hfill
\\
\begin{subfigure}{0.32\textwidth}
    \centering
    \includegraphics[width=0.99\textwidth]{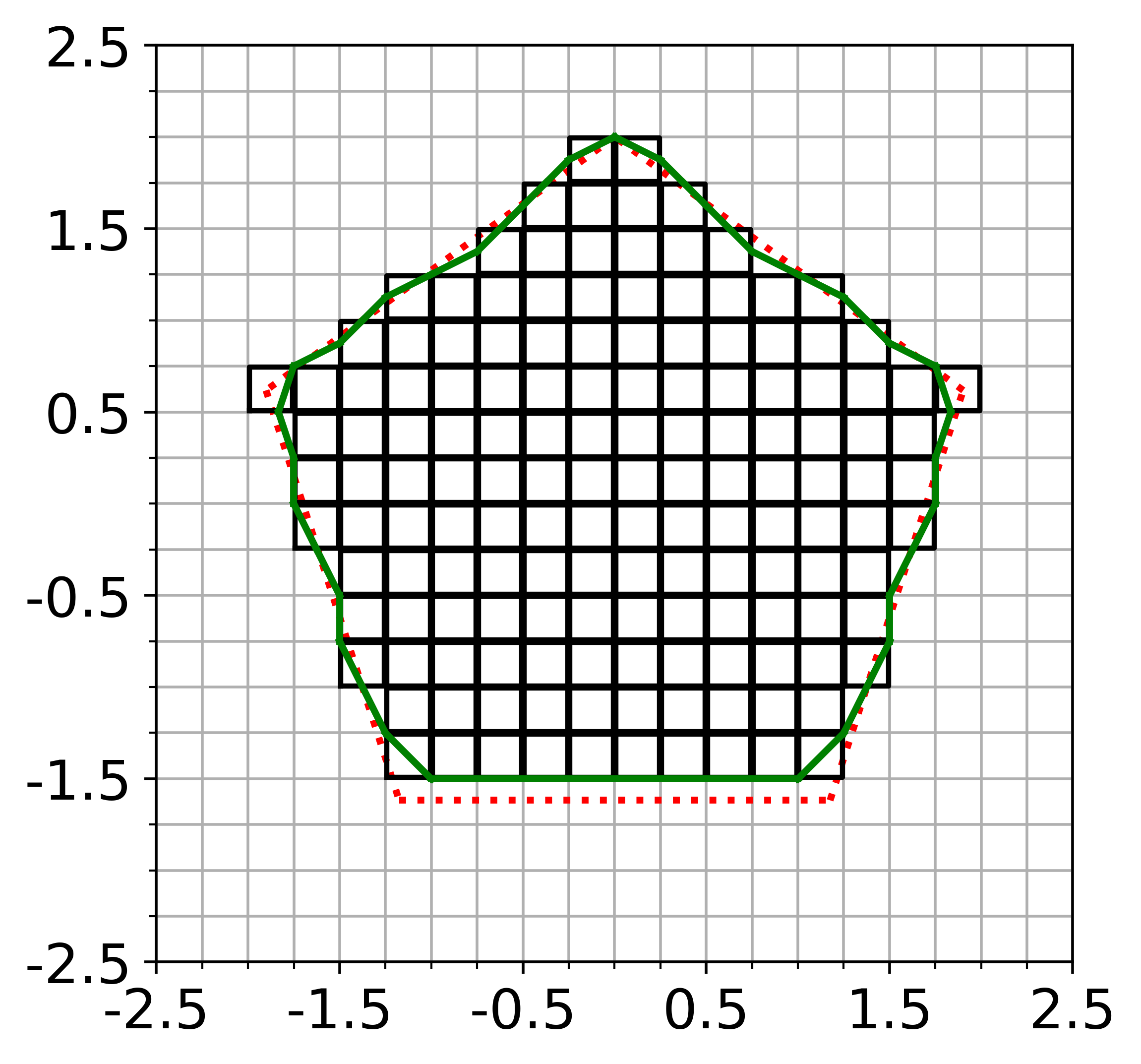}
    \caption{Pentagon.}\label{mm_pentagon}
\end{subfigure}
\begin{subfigure}{0.32\textwidth}
    \centering
    \includegraphics[width=0.99\textwidth]{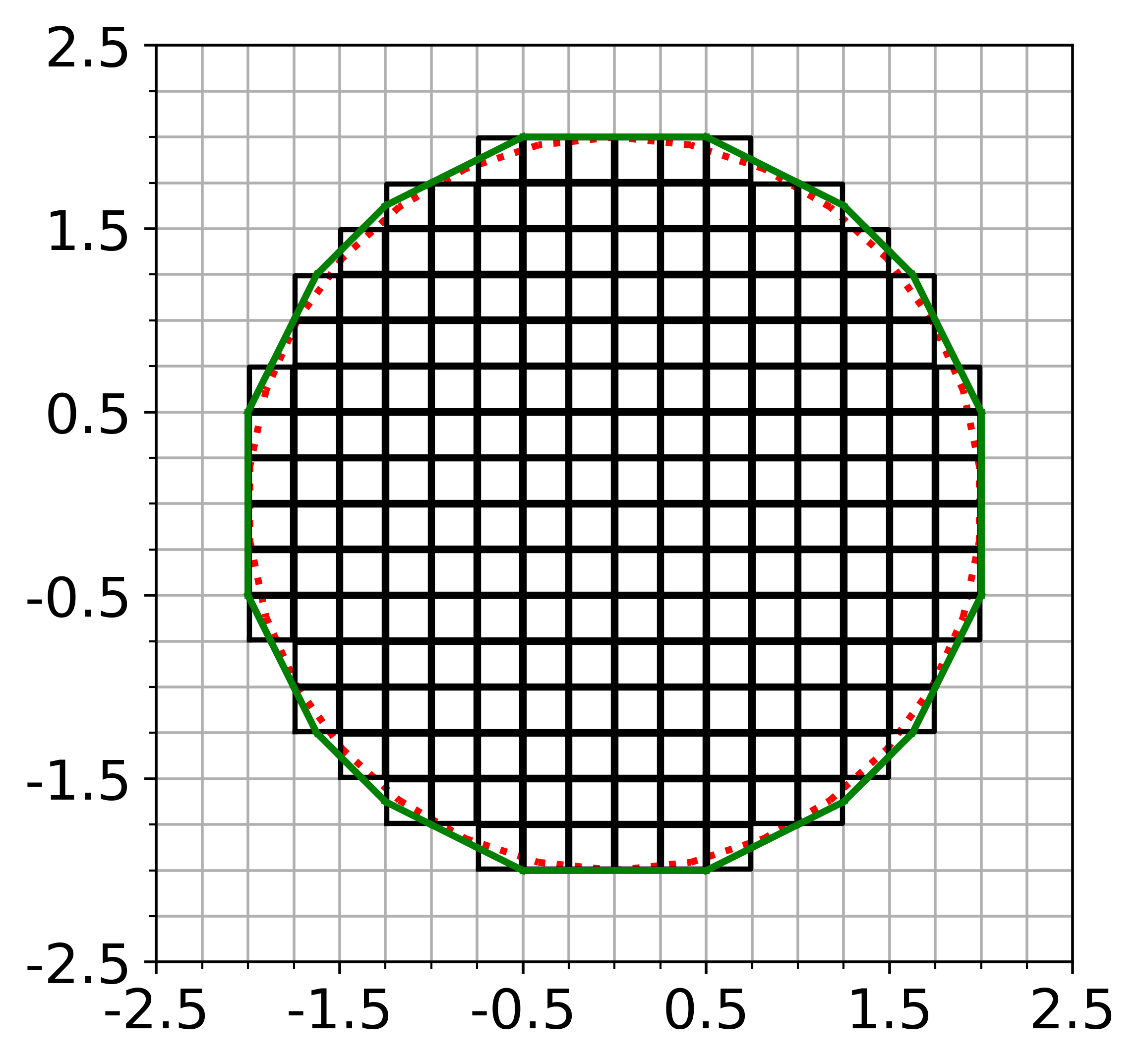}
    \caption{Triacontagon.}\label{mm_circle}
\end{subfigure}
\caption{Example surfaces produced by motion mapping. The black squares are voxels, and the green lines are the generated boundary surface. The red dotted line is the ideal regular polygon which was used to generate the voxelized structure. Voxel ratio and grid ratio are both eight.} 
\label{mm_geom}
\end{figure}

To assess the quality of surfaces generated in the motion mapping module of marching windows, several shapes are approximated with voxelized structures and surfaces are generated from these structures. The five convex polygons used in this analysis are illustrated in Fig.~\ref{mm_geom}: a triangle, a square, a diamond, a pentagon, and a 30-sided polygon (triacontagon). These five shapes are analyzed in order to understand the effectiveness of marching windows on various geometric features that can be present in larger structures. For example, when representing mechanical parts, there can be curved geometric features on rolling components that can be approximated with a portion of a circular geometry. Other sections of the geometry like gear teeth can have sharp corners. The shapes presented in this section are chosen to assess the capabilities of marching windows for these different geometric structures. The triangle, square, and pentagon are analyzed to estimate the errors generated in geometries with sharp corners at varying internal angles. The diamond is used to assess the ability of marching windows to capture shape edges that are at an angle with respect to the voxel edges. The triacontagon is used as a numerical estimate of a circle. All shapes are regular polygons with a circumradius of two, where the units are arbitrary. The edges of the ideal shapes are shown in red in Fig.~\ref{mm_geom}.

These shapes are used to generate the voxelized structures shown in gray. The centroid of each voxel created in this structure is located inside or on the ideal shape borders. Voxel side lengths are 0.25, corresponding to a voxel ratio of 
$v_r = 8$.
These voxels are then weighted, and their volumes are distributed to marching windows grid nodes according to the procedure outlined above. The cell size in the marching windows grid is also 0.25, corresponding to a grid ratio of
$g_r = 8$.
The marching windows grid, the voxelized structure, and the ideal shape are all centered at the origin.
Once weighted voxel volumes are distributed to nodes, the marching squares procedure generates the green boundary surfaces shown in Fig.~\ref{mm_geom}. In general, the geometric features of each of the voxelized structures are captured by the corresponding surface. For example, the curvature of the generated circle surface closely follows the ideal circle. The angles between the edges of the other four shapes are also captured by the generated surfaces.

One notable feature of these boundary surfaces is their tendency to round off and retract sharp corners in the voxelized geometry. This effect is illustrated on the bottom corners of the triangle, all corners of the square, and the left and right corners of the pentagon. Marching squares, while being a robust method to generate closed and non-intersecting surfaces, cannot reliably resolve features like sharp corners. Any feature on a geometry which is similar in size to or smaller than a grid cell has the possibility of being rounded off when marching squares is applied to generate a surface. The circle in Fig.~\ref{mm_circle} is well-approximated by the marching squares boundary surface. The surface does have straight lines where the edge voxels form straight lines, but overall, the curvature of the surface closely hugs both the voxelized structure edge and the ideal circular geometry. The circular geometry does not have small features relative to the grid cell size, so the round-off effect is not as much of a concern as for sharp-cornered geometries.

In addition to feature round-off, some surface jaggedness is also possible. While the square and diamond geometries in Fig.~\ref{mm_geom} have smooth boundary surfaces far from the corners, the pentagon and triangle edges are more jagged. Note that the bottom edges of both the triangle and pentagon are smooth like the edges of the square. For both of these bottom edges and all the edges of the square, there is a single, flat layer of edge voxels that the voxel weighting and marching squares procedures can represent perfectly. The diamond has smooth edges as well despite not having flat voxelized edges. Since the diamond edges are at a consistent 45$\degree$ incline relative to the $x$-axis, the voxelized structure can represent the edges in a uniform way. For example, the top-left edge of the diamond has a rise-over-run slope of $1/1$, so the edge voxels in the voxelized structure can represent the slope by having each voxel one space up and one space to the right of the previous one. This consistency allows marching squares to generate surface elements with a consistent slope.
On the other hand, the pentagon and triangle have slopes that are not at an angle relative to the $x$-axis that is a whole-number multiple of 45$\degree$. As a result, the voxelized structure has an inconsistent slope. For example, the right-bottom edge of the ideal pentagon rises at a 72$\degree$ angle relative to the $x$-axis. The voxelized structure is made up of elements of fixed size and shape, so angles which are not a whole-number multiple of 45$\degree$ can only be represented at length scales larger than one voxel. The right-bottom edge of the pentagon is represented by the voxels with a rise of three voxels with every one voxel running to the right, whereas the ideal 72$\degree$ slope has an approximate $3.1/1$ rise-over-run. In short, that edge of the voxelized pentagon is a series of flat walls with a height of three voxels each. The result of this inconsistent slope is that the surface generated by marching squares has relatively drastic changes in slope between elements.

\begin{figure}[ht!]
\centering
\begin{subfigure}{0.32\textwidth}
    \centering
    \includegraphics[width=0.99\textwidth]{geometries/pent0_geom_8.png}
    \caption{$g_r = 8$.}\label{pent_8}
\end{subfigure}
\hfill
\begin{subfigure}{0.32\textwidth}
    \centering
    \includegraphics[width=0.99\textwidth]{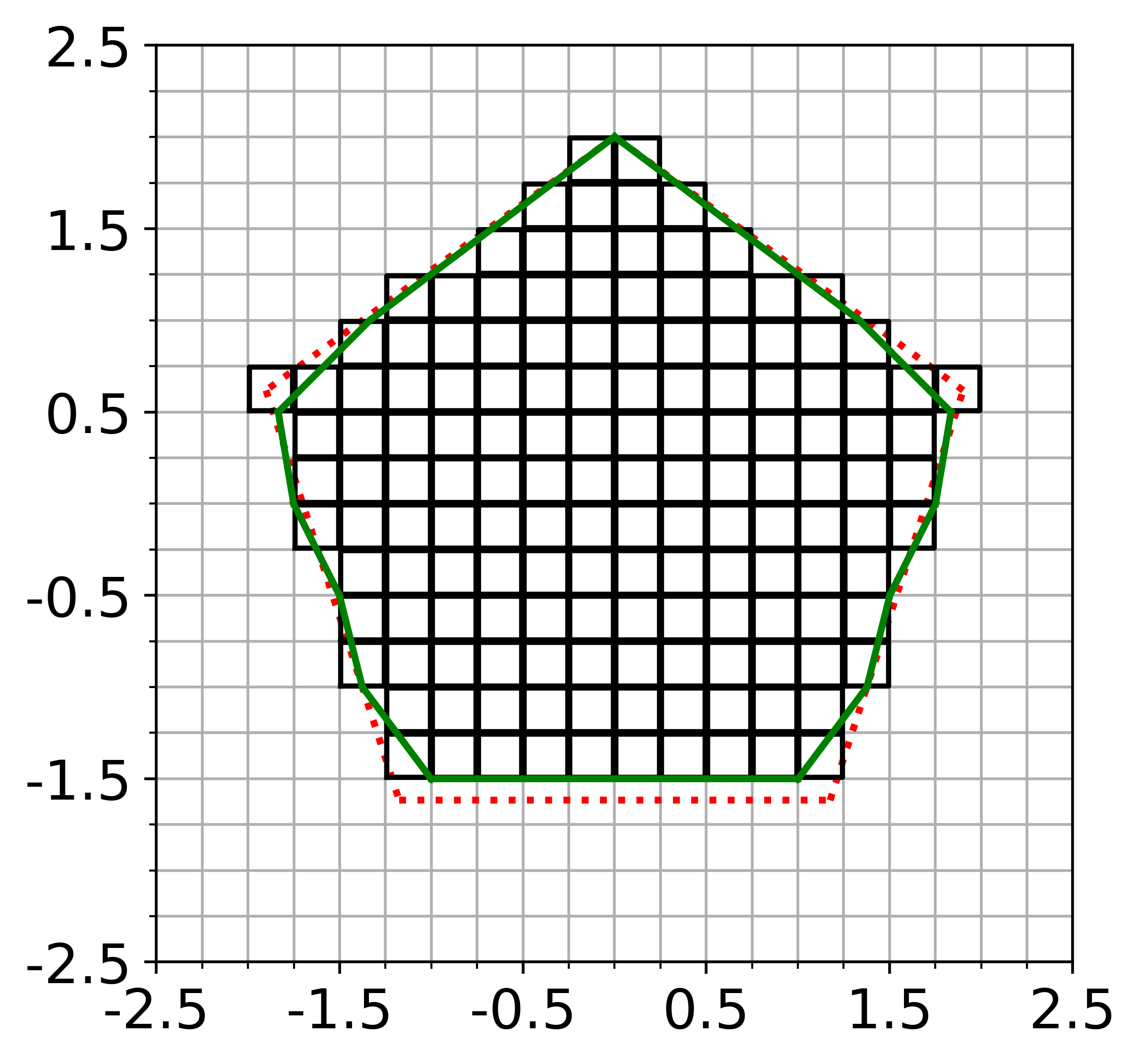}
    \caption{$g_r = 4$.}\label{pent_4}
\end{subfigure}
\hfill
\begin{subfigure}{0.32\textwidth}
    \centering
    \includegraphics[width=0.99\textwidth]{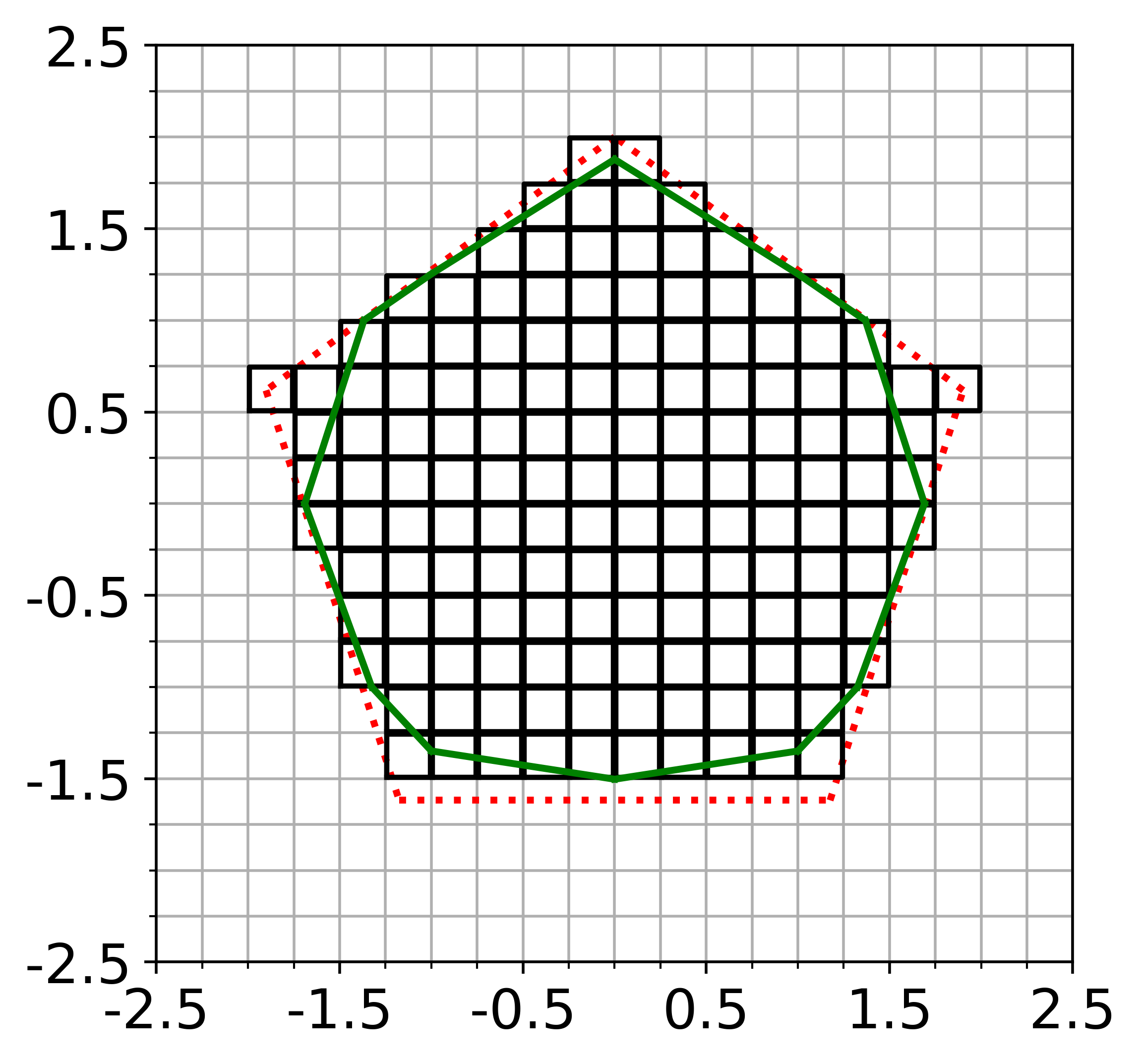}
    \caption{$g_r = 2$.}\label{pent_2}
\end{subfigure}
\hfill
\caption{Motion mapping for a voxelized pentagon at varying grid ratios. Voxel ratio is eight in all cases.} 
\label{pent}
\end{figure}

Surface jaggedness can be mitigated by using larger marching windows grid cells, i.e. a smaller grid ratio. The same voxelized pentagon from Fig.~\ref{mm_pentagon} is shown in Fig.~\ref{pent} with generated surfaces at varying grid ratios. The surface in Fig.~\ref{pent_8} is the same as in Fig.~\ref{mm_pentagon} with a grid ratio of $g_r = 8$. Reducing grid ratio by half to $g_r = 4$ results in grid cells of twice the size. The resulting boundary surface is shown in Fig.~\ref{pent_4}. Some jaggedness is still visible on the bottom-left and bottom-right edges, but the top two edges are substantially smoother than in Fig.~\ref{pent_8}. On the other hand, using a small grid ratio can also worsen the corner round-off described above. A surface generated with a grid ratio of $g_r = 2$ is shown in Fig.~\ref{pent_2}. Compared to the surface in Fig.~\ref{pent_4}, all corners of the pentagon have visibly retracted inwards, especially the left and right corners. 

The flexibility of marching windows to allow differing resolutions of voxels and grid cells makes it suitable for coupling simulation methods with differing resolution requirements. The voxel resolution is independent of grid resolution, so the grid resolution can be chosen to suit the needs of the fluid simulation rather than chosen to conform to the needs of the solid simulation.

As mentioned above, voxelized structures do not have a well-defined surface. The lack of an accurate reference surface poses a difficulty in specifying an error metric for the quality of a marching windows surface. The error metric adopted in this article is based on the assumption that any perfect surface that could be defined for a voxelized structure should have all voxel centroids inside of it. Similarly, a perfect surface should not contain an abundance of empty volume with no voxels. For example, the empty voxel spaces, or voxel voids, in Fig.~\ref{mm_geom} are the squares outside the voxelized geometries whose edges are light gray. While the boundary surface may penetrate into these voxel voids at certain points, a perfect surface should not contain their centroids. In short, a perfect boundary surface of a voxelized structure should contain all voxel centroids and contain none of the void centroids. If these two centroid containment conditions are satisfied by a surface generated by marching squares, then all features of the voxelized geometry should be represented by the surface. In theory, a surface could be generated that satisfies the containment conditions while still including thin, spurious features that snake between voxel centroids or void centroids. However, as mentioned above, marching squares has a tendency to clip very small features rather than spuriously generate them.

\begin{figure}[ht!]
\centering
\begin{subfigure}{0.45\textwidth}
    \centering
    \includegraphics[width=0.99\textwidth]{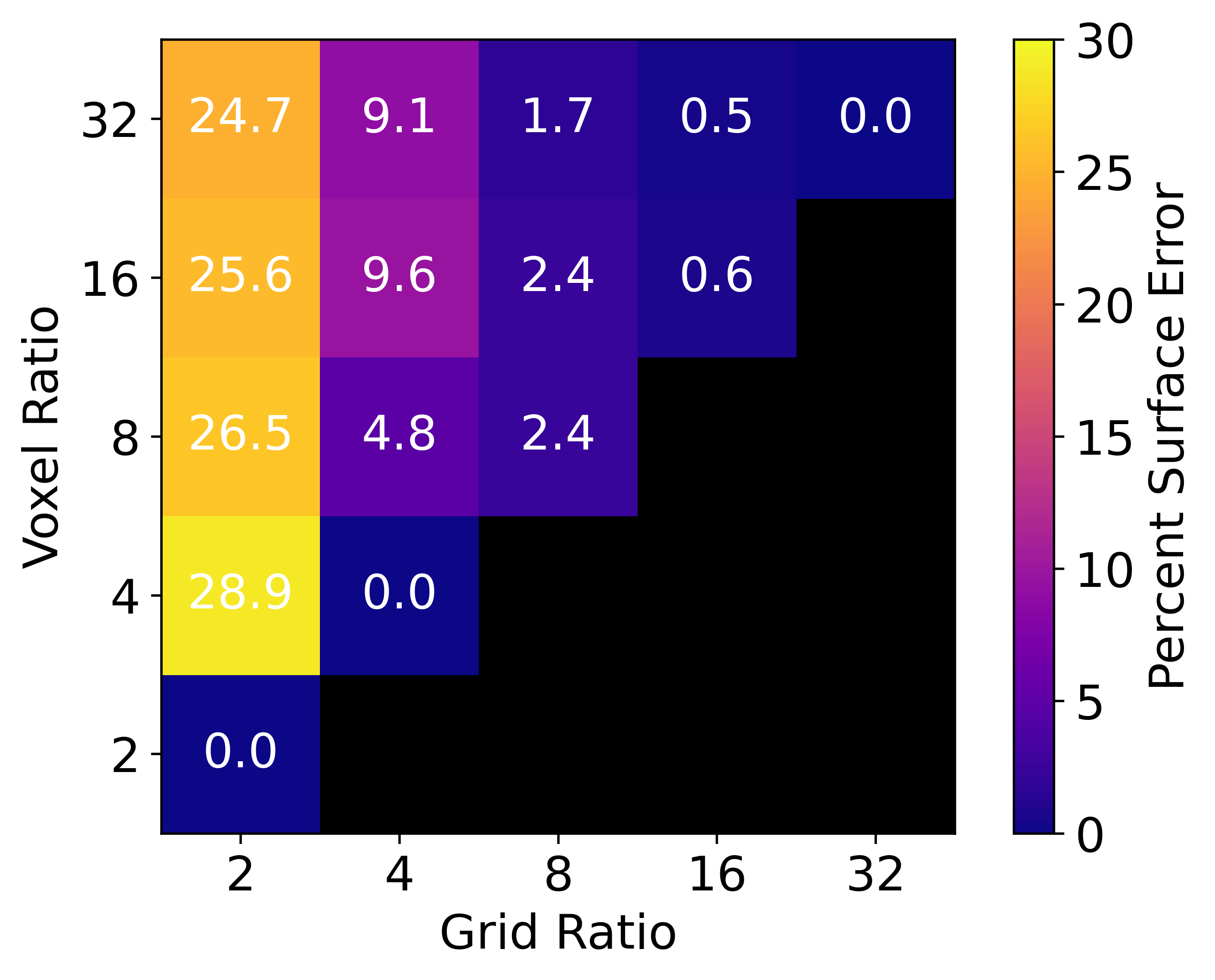}
    \caption{Triangle.}\label{mheat_triangle}
\end{subfigure}
\hfill
\begin{subfigure}{0.45\textwidth}
    \centering
    \includegraphics[width=0.99\textwidth]{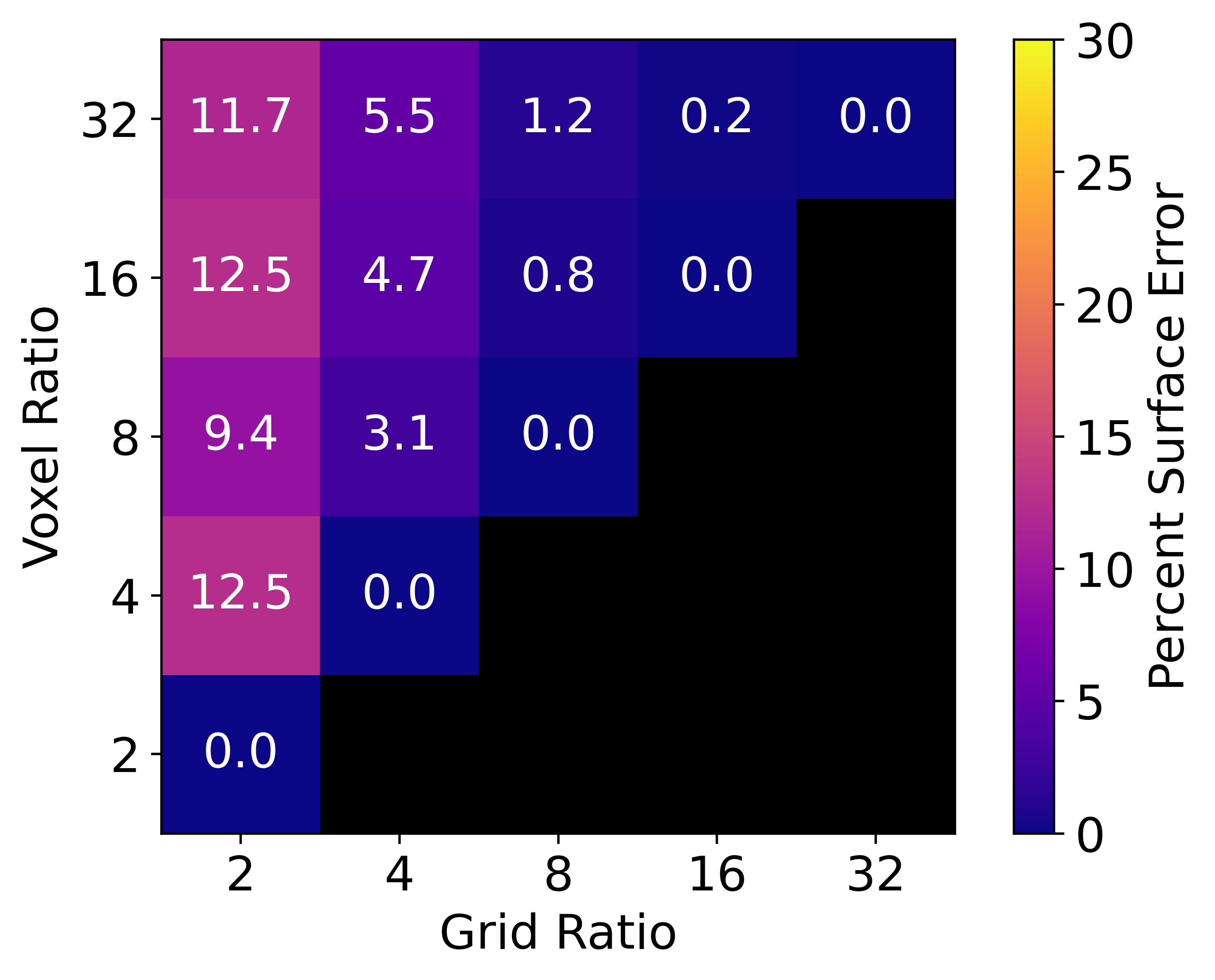}
    \caption{Square.}\label{mheat_square}
\end{subfigure}
\hfill
\begin{subfigure}{0.45\textwidth}
    \centering
    \includegraphics[width=0.99\textwidth]{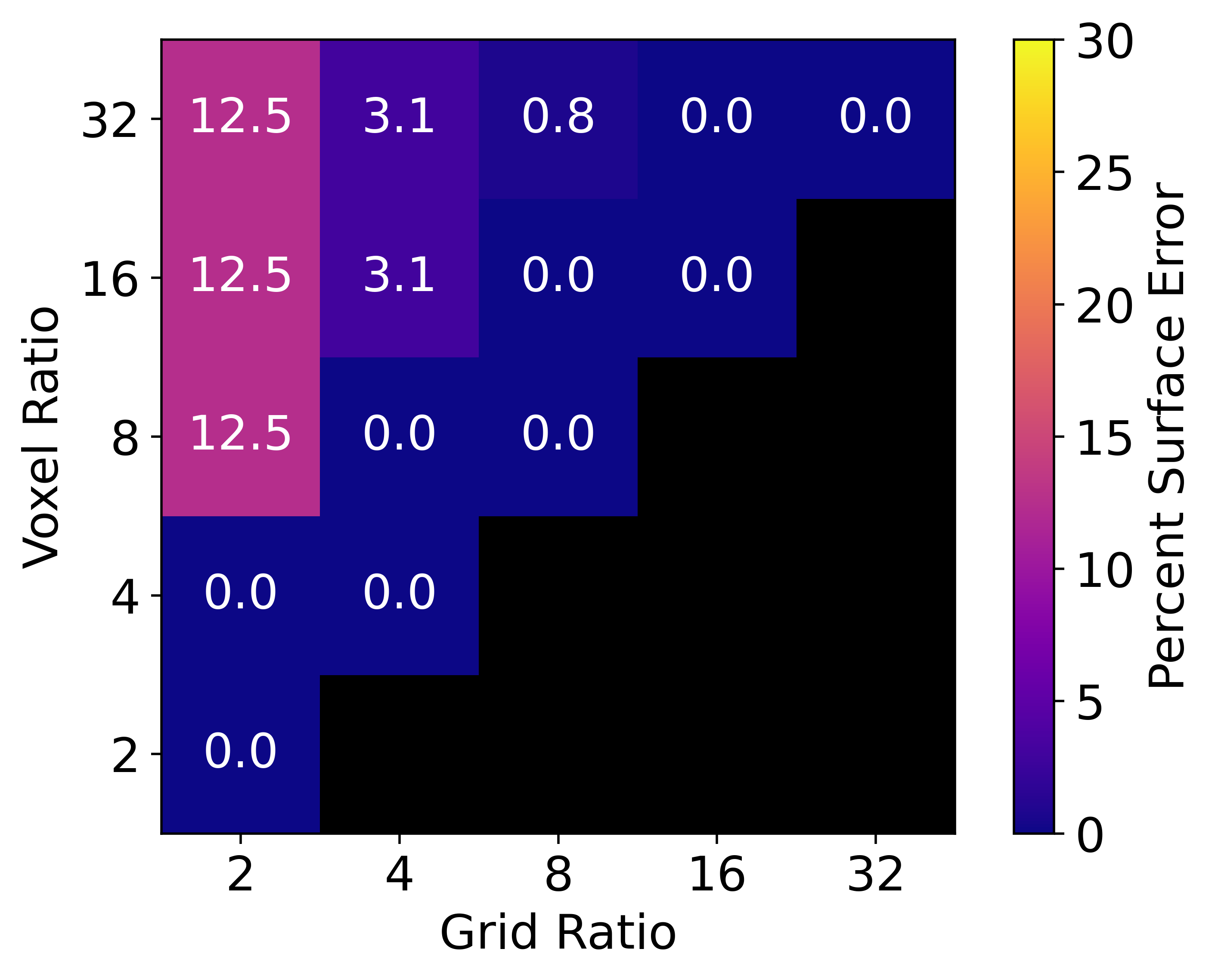}
    \caption{Diamond.}\label{mheat_diamond}
\end{subfigure}
\hfill
\begin{subfigure}{0.45\textwidth}
    \centering
    \includegraphics[width=0.99\textwidth]{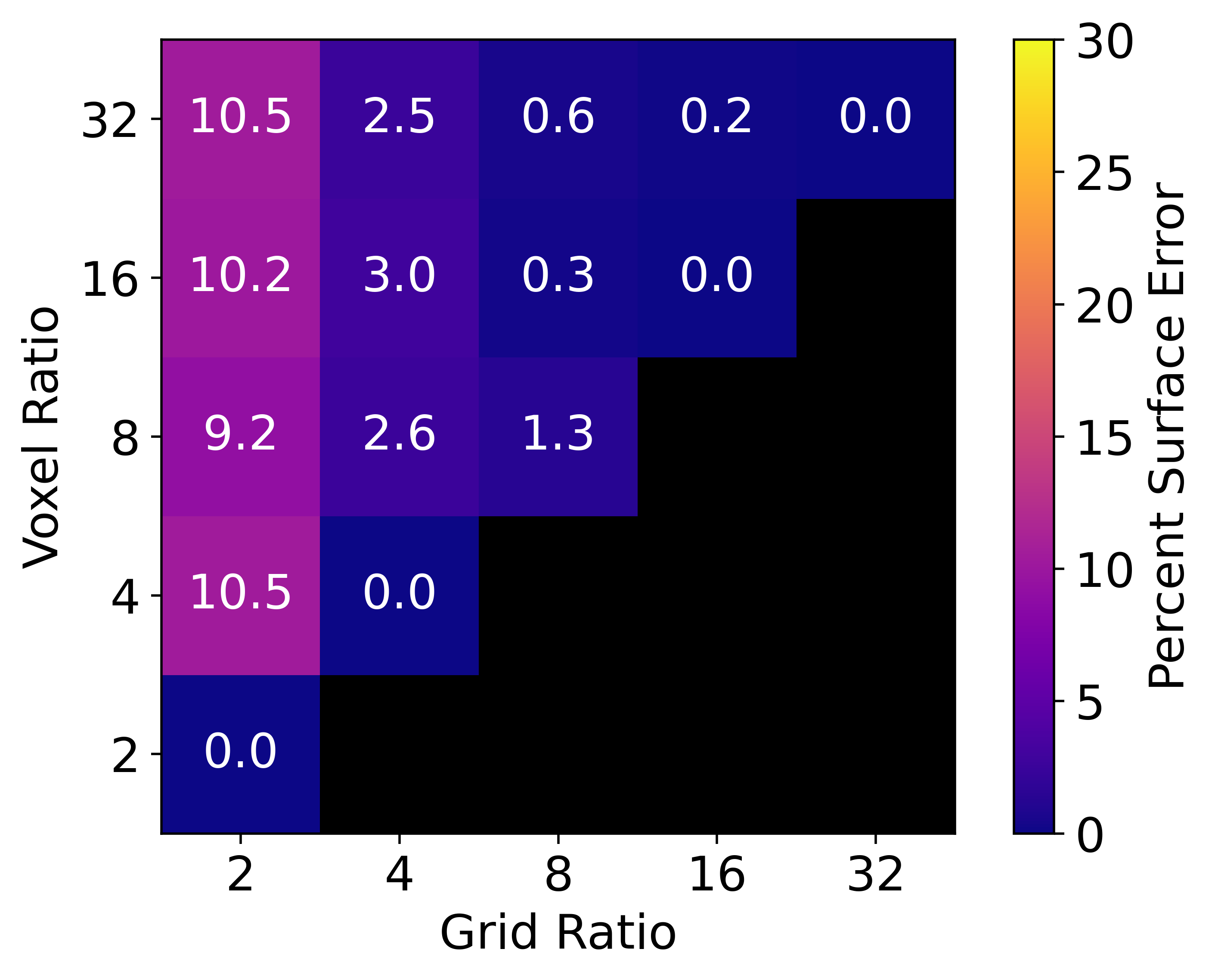}
    \caption{Pentagon.}\label{mheat_pentagon}
\end{subfigure}
\hfill
\begin{subfigure}{0.45\textwidth}
    \centering
    \includegraphics[width=0.99\textwidth]{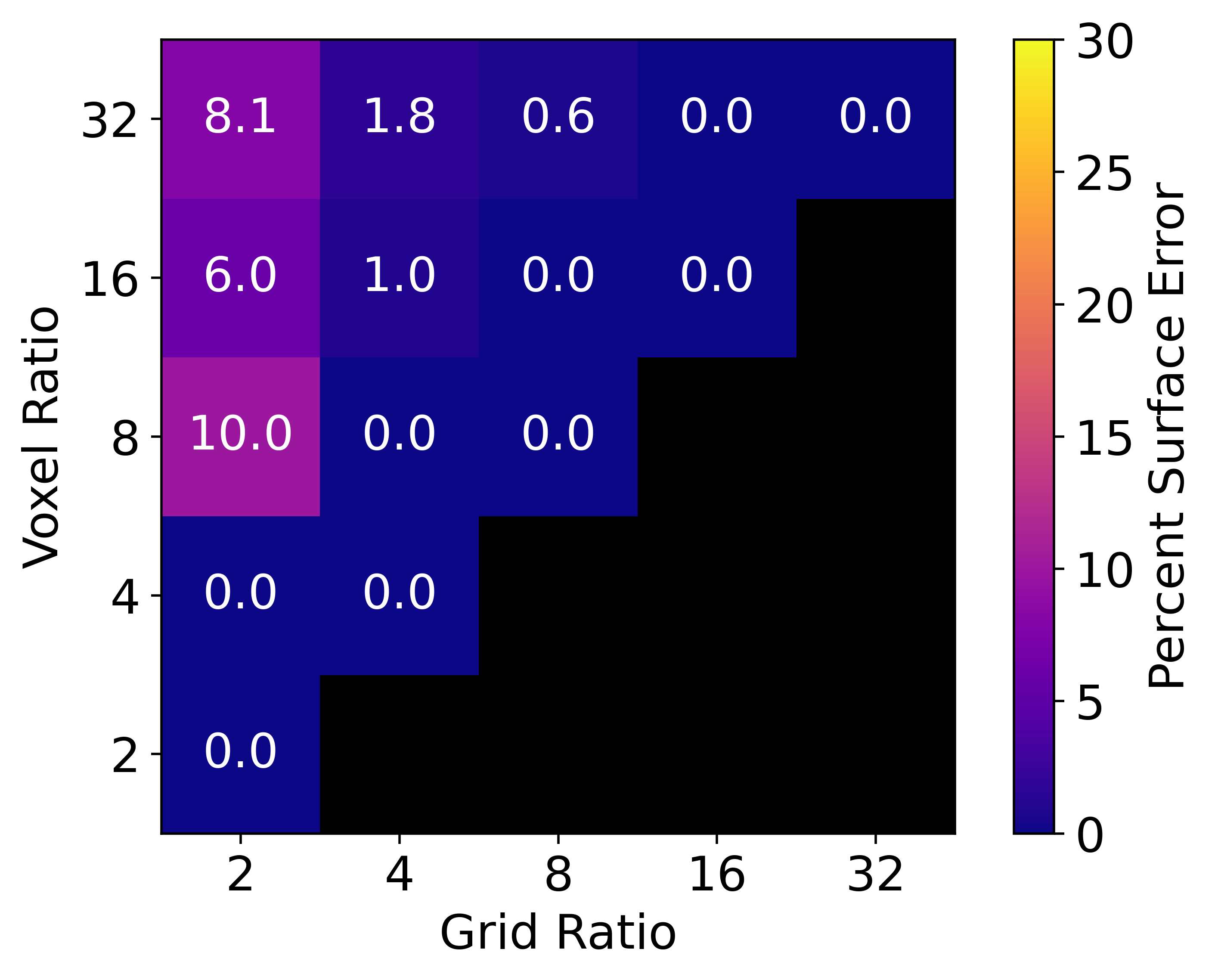}
    \caption{Triacontagon.}\label{mheat_circle}
\end{subfigure}
\hfill
\caption{Centroid containment error in motion mapping for regular polygons at varying voxel ratios and grid ratios. The same shapes are used as in Fig.~\ref{mm_geom}. Error is quantified by total area of voxels whose centroids lie outside the generated boundary surface, added to the area of voxel voids whose centroids lie inside the surface. This summed area is expressed as a percentage of the ideal polygon area.} 
\label{mm_heat}
\end{figure}

Each of the ideal shapes from Fig.~\ref{mm_geom} were used to generate a voxelized structure and create a boundary surface at varying voxel ratios and grid ratios. A wide range of voxel sizes and cell sizes was used to determine the necessary voxel ratios and grid ratios to generate an accurate surface for each shape. The surface quality of each of these mappings was estimated based on the number of voxels and voxel voids which violated the centroid containment conditions, and the results are shown in Fig.~\ref{mm_heat}. The total number of violating voxels ($n_{\text{vox},\text{err}}$) and violating voxel voids ($n_{\text{void},\text{err}}$) is used to calculate containment error ($E_\text{mm}$). The area of these violating voxels and voids is expressed as a percentage of the area of the ideal polygon ($A_\text{poly}$),
\begin{equation}
    E_{\text{mm}} = \frac{(n_{\text{vox},\text{err}} + n_{\text{void},\text{err}})A_\text{vox}}{A_\text{poly}} \cdot 100\%
    \label{eq_mm_err}
\end{equation}
where $A_\text{vox}$ is the area of an individual voxel. The containment error is indicated as a percentage in Fig.~\ref{mm_heat} for each grid ratio and voxel ratio combination. The blacked-out combinations in the heatmaps have a grid ratio which is larger than the voxel ratio, so surfaces are not generated for them.

The results in Fig.~\ref{mm_heat} have some similarity and some difference across the shapes. For all shapes, centroid containment errors are less than 2.5\% for all combinations with grid ratios $g_r \geq 8$. All shapes also have little to no containment error when $g_r = v_r$, even at low ratio values. However, the combination $g_r = v_r = 2$ having an error $E_\text{mm} = 0.0\%$ for all shapes is likely partially a result of having very few, very large voxels that are easily contained even by poor surfaces.
Of all the analyzed shapes, the circle is the most robust to varying voxel and grid ratios. There are no sharp features for the motion mapping to capture, and all results at grid ratios $g_r \geq 4$ show less than 2\% containment error.
The triangle is the most sensitive shape to the choice of voxel and grid ratios, as it has the most extremely acute angles of the analyzed geometries. However, even the triangle shows less than 2.5\% error at grid ratios $g_r \geq 8$. The diamond, square, and pentagon produce results in between the extremes of the triangle and circle. Even though the edges of the diamond are at a 45$\degree$ angle to the voxel alignment, the containment errors are generally comparable to or better than the corresponding square containment errors. The pentagon, despite the jaggedness described earlier, shows similar containment errors to the square and diamond at corresponding ratio combinations.

Based on the results of Fig.~\ref{mm_heat}, geometric features similar to the polygons tested can be captured by the motion mapping procedures described in this article for grid ratios of $g_r \geq 8$.

\FloatBarrier

\section{Flux Mapping}
\label{sec_fluxmap}

\subsection{Flux Mapping Method}
The flux mapping module of marching windows relates the surface elements generated in the motion mapping module to voxels in the original structure. The relation of voxels to surfaces allows the conversion of surface quantities to voxel quantities. Depending on the application, possible surface quantities include heat loads, surface reactions, and mechanical forces. These quantities will generically be referred to as surface scalars. Flux mapping is performed in two steps. Each step is described below along with details on 3D implementation of flux mapping. The first step is determining the visibility of edge voxels to surface elements. The exposed faces of these edge voxels are projected onto nearby surface elements to calculate the overlapping projected area. This area is the length of the overlap of the surface element and the projected face, multiplied by unit depth. The second step is the distribution of the surface scalars to voxels. For each surface element, the scalar is divided proportionally among the voxels according to the overlapping projected areas calculated in the first step.

\subsubsection{Face Projection}
In 2D structures, each voxel has four edges as its faces. Any of the exposed faces can potentially project onto a surface element. Each of the exposed faces ($f$) of each edge voxel ($v$) are projected onto nearby surface elements. If the projection of the face overlaps with the surface element ($t$), then the length of this overlap ($L_{\text{clip},f,t}$) is used in the second flux mapping step to calculate the share of the scalar quantity that is received by voxel face $f$ from surface element $t$.

\begin{figure}[ht!]
\centering
    \includegraphics[width=0.72\textwidth, clip]{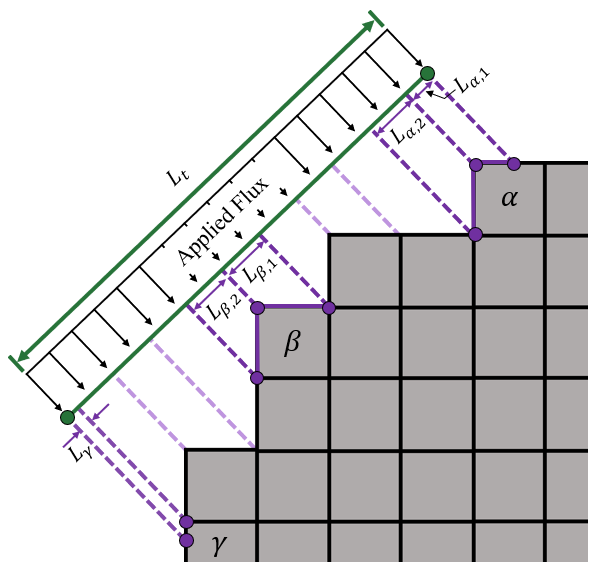}
\hfill
\caption{A surface element (green) distributing its applied scalar flux to voxels (gray) by face projection.} 
\label{scalar_app}
\end{figure}

Figure~\ref{scalar_app} illustrates the process of projecting exposed voxel faces onto surface elements. The surface element (green) of length $L_t$ has a surface scalar that must be distributed to nearby voxels. The flux mapping process models this surface-to-voxel transfer like a uniform flux applied in the direction of the surface normal. The transfer is akin to a window passing light from the outside to interior surfaces. In Fig.~\ref{scalar_app}, all of the voxel faces that partially or fully overlap the surface element when projected onto it are indicated with purple lines extending from the face to the element. The two exposed faces of voxel $\beta$, when projected, have lengths of $L_{\beta,1}$ and $L_{\beta,2}$ for the top and left face, respectively. The rest of the exposed faces of edge voxels between voxel $\alpha$ and voxel $\gamma$ behave similarly to voxel $\beta$ in this projection step.
The left face of voxel $\alpha$ also has a projected length $L_{\alpha,2}$. The top face of voxel $\alpha$ and the left face of voxel $\gamma$ partially project onto the surface element, but their projections extend past the limits of the surface element. Only the portions of the projected lengths that intersect the surface element ($L_{\alpha,1}$ and $L_{\gamma}$) are considered during surface scalar distribution to voxels. In a 3D version of flux mapping, the surface elements would be triangles and the voxel faces would be squares rather than both surface elements and voxel faces being line segments. The projection of the square faces onto the surface triangles would work in the same way as in 2D, except that overlap areas would be calculated as an intersection of polygons rather than an intersection of line segments with unit depth.

\subsubsection{Scalar Transfer}

The scalar of each surface element ($F_t$) is divided between all voxel faces that have a nonzero overlap length. The fraction of $F_t$ received by each face ($F_{v,f,t}$) is weighted by the overlap length calculated in step one,
\begin{equation}
\label{eq_areafrac}
    F_{v,f,t} = F_t \frac{L_{\text{clip},f,t}}{\sum_i^{n_{f,t}} L_{\text{clip},i,t}}
\end{equation}
where $L_{\text{clip},i,t}$ is the overlap length of the $i$-th voxel face of $n_{f,t}$ total voxel faces projected onto element $t$. Using Eq.~\ref{eq_areafrac}, the total scalar value $F_t$ is guaranteed to be conserved when distributed to voxel faces. This conservation of total scalar value is only violated in the case of a surface element having zero projected overlap with any voxel face. In theory, a surface element with no projected face overlap is possible, but this condition should be rare or nonexistent for accurate surfaces generated by the motion mapping module.

Voxel faces can be projected onto multiple surface elements. The total scalar quantity of each voxel face ($F_{v,f}$) is the sum of the contributions from all $n_\text{tri}$ surface elements,
\begin{equation}
    F_{v,f} = \sum_t^{n_\text{tri}} F_{v,f,t}
\end{equation}
Likewise, the total scalar quantity of each voxel ($F_v$) is a sum over the scalars of all of its exposed faces. For a voxel $v$ with $n_{\text{exp},v}$ exposed faces, its total scalar is
\begin{equation}
    F_v = \sum_f^{n_{\text{exp},v}} F_{v,f}
\end{equation}
The scalar transfer process described above is extendable to vector quantities by treating all vector components as separate scalars. Each surface vector component is distributed to the voxels independently and uses the same exposed face projections to do so.

In the example geometry of Fig.~\ref{scalar_app}, the top face of voxel $\beta$ receives a scalar load in accordance with Eq.~\ref{eq_areafrac} as
\begin{equation}
    F_{\beta,1,t} = F_t \frac{L_{\beta,1}}{L_t}
\end{equation}
The scalar load of the left face of voxel $\beta$ is calculated similarly, so the total scalar load on voxel $\beta$ is
\begin{equation}
    F_{\beta} = F_t \left(\frac{L_{\beta,1} + L_{\beta,2}}{L_t}\right)
\end{equation}
If the faces of voxel $\beta$ had projected onto any other surface elements, the contributions of those elements to voxel $\beta$ would be added to $F_{\beta}$. Voxels $\alpha$ and $\gamma$ and the rest of the edge voxels between them receive their portions of the surface scalar in the same way. Voxels beyond $\alpha$ and $\gamma$ have zero projected area on the surface element and thus receive no portion of its scalar quantity. In 3D flux mapping, the only necessary change to the scalar distribution procedure would be to use the projected overlap areas in Eq.~\ref{eq_areafrac} instead of the overlap lengths.

\subsection{Error Analysis}

The polygonal geometries from Sec.~\ref{sec_motionmap} are analyzed to assess the error of the marching windows flux mapping module. An analytical test flux function is applied to each surface element based on angular position, and the resulting surface scalars are transferred to the voxels through flux mapping. The voxel scalars are then converted into equivalent fluxes. Each voxel flux is compared to the analytical flux at the position of the voxel to determine the error generated in the flux mapping procedure.

Motion mapping is used on the same five geometries illustrated in Fig.~\ref{mm_geom} to generate a boundary surface. Voxel ratio and grid ratio are both eight. Once the surface is created, a dummy flux ($\gamma_t$) is applied to each surface element ($t$),
\begin{equation}
    \label{eq_gamma}
    \gamma_t = \cos\theta_{c,t} + 2
\end{equation}
where $\theta_{c,t}$ is the angle of the surface element midpoint relative to the positive $x$-axis. A sinusoidal function was chosen so that the variation in flux with angular position would be graphically visible. The flux $\gamma_t$ corresponds to a surface scalar $\gamma_t L_t$ for a surface element of length $L_t$.
The flux mapping procedure then distributes the surface scalars of each element to the voxels, and the resulting voxel scalars are divided by a characteristic voxel length to obtain an equivalent flux. The characteristic length used for each voxel is the sum of all the projected lengths of all its exposed faces. The voxel flux is then compared against the analytical flux calculated at the angular coordinate of the voxel centroid.

\begin{figure}[ht!]
\centering
\begin{subfigure}{0.49\textwidth}
    \centering
    \includegraphics[width=0.99\textwidth]{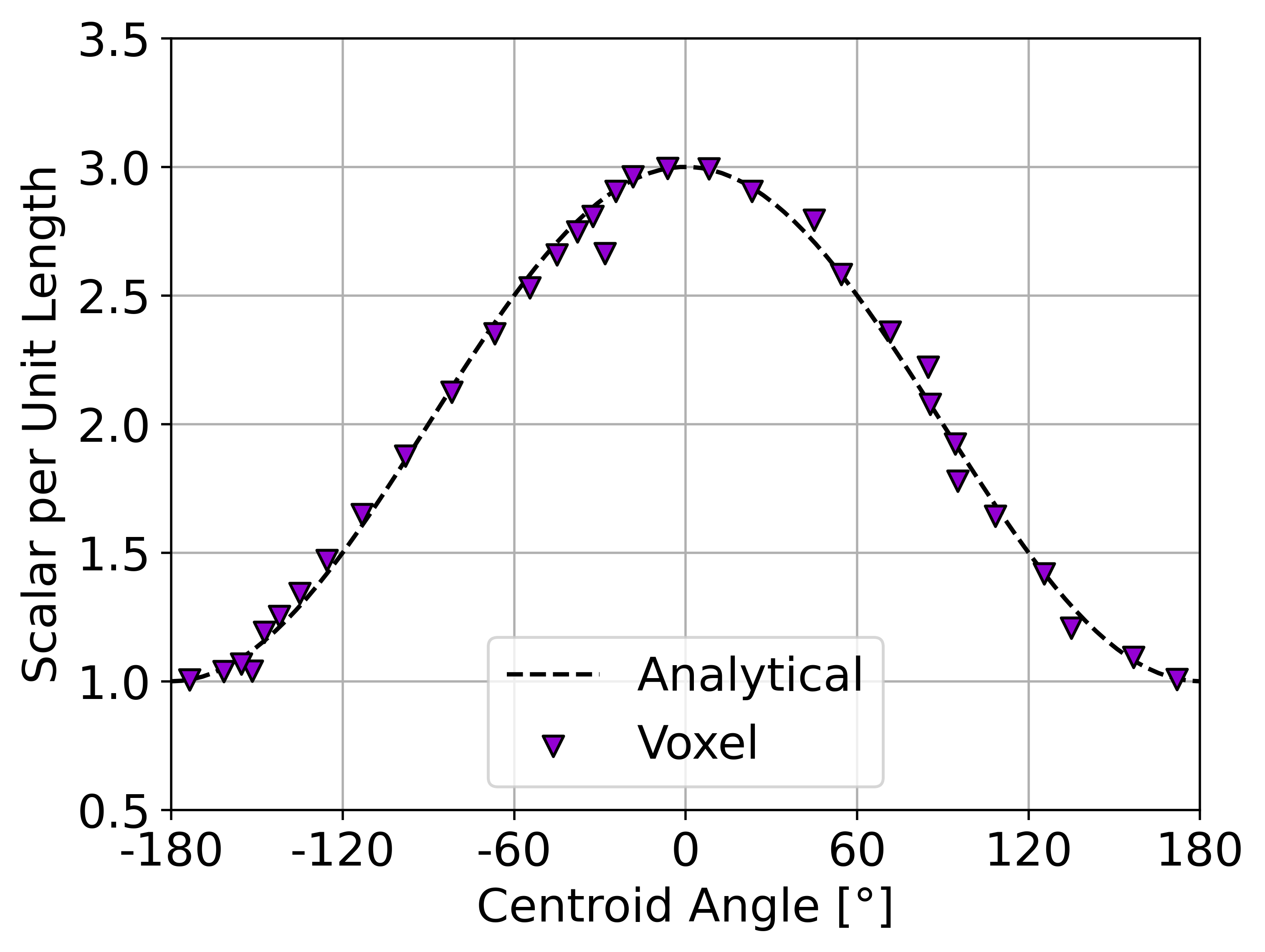}
    \caption{Triangle.}\label{fm_triangle}
\end{subfigure}
\hfill
\begin{subfigure}{0.49\textwidth}
    \centering
    \includegraphics[width=0.99\textwidth]{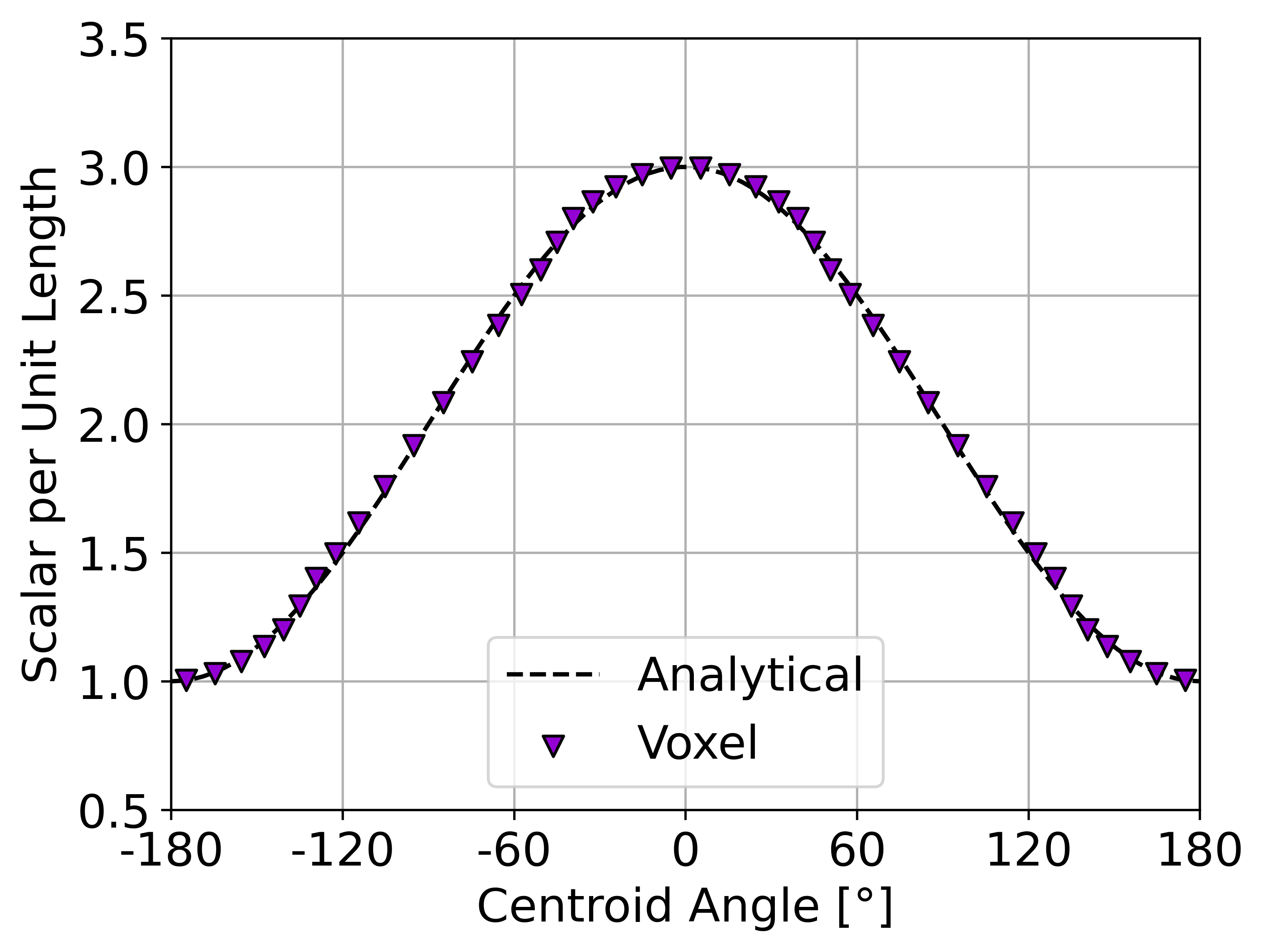}
    \caption{Square.}\label{fm_square}
\end{subfigure}
\hfill
\begin{subfigure}{0.49\textwidth}
    \centering
    \includegraphics[width=0.99\textwidth]{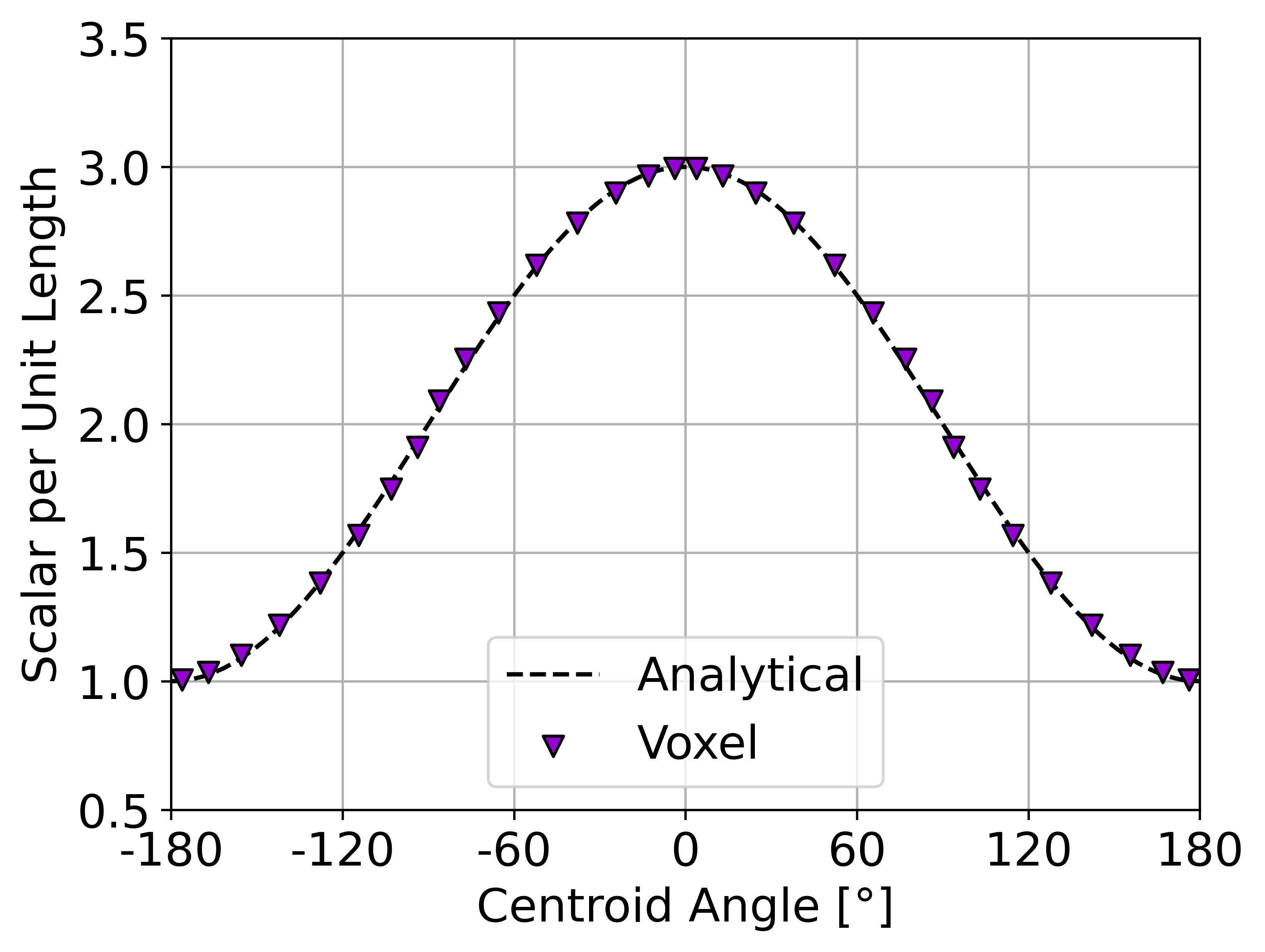}
    \caption{Diamond.}\label{fm_diamond}
\end{subfigure}
\hfill
\begin{subfigure}{0.49\textwidth}
    \centering
    \includegraphics[width=0.99\textwidth]{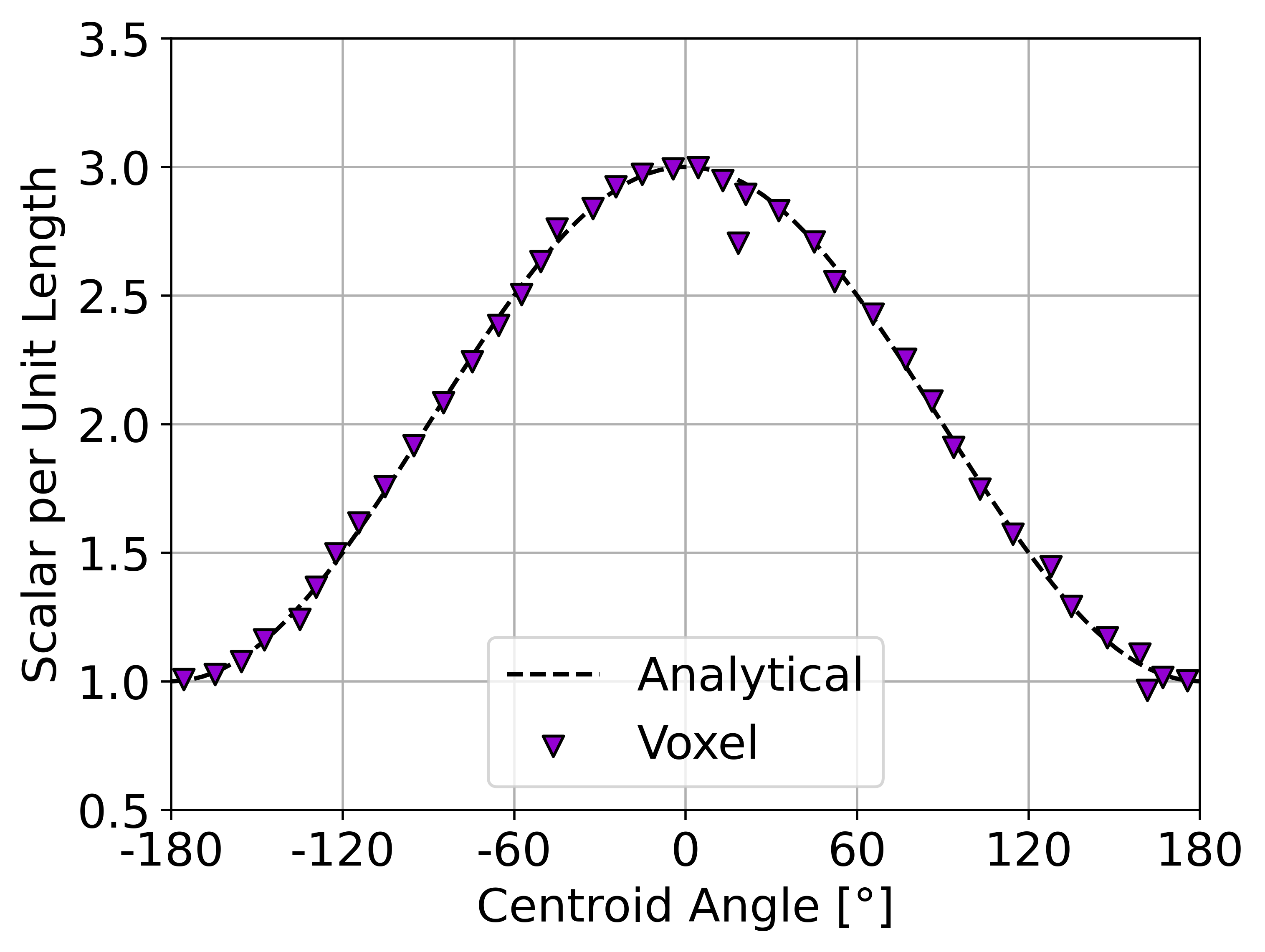}
    \caption{Pentagon.}\label{fm_pentagon}
\end{subfigure}
\begin{subfigure}{0.49\textwidth}
    \centering
    \includegraphics[width=0.99\textwidth]{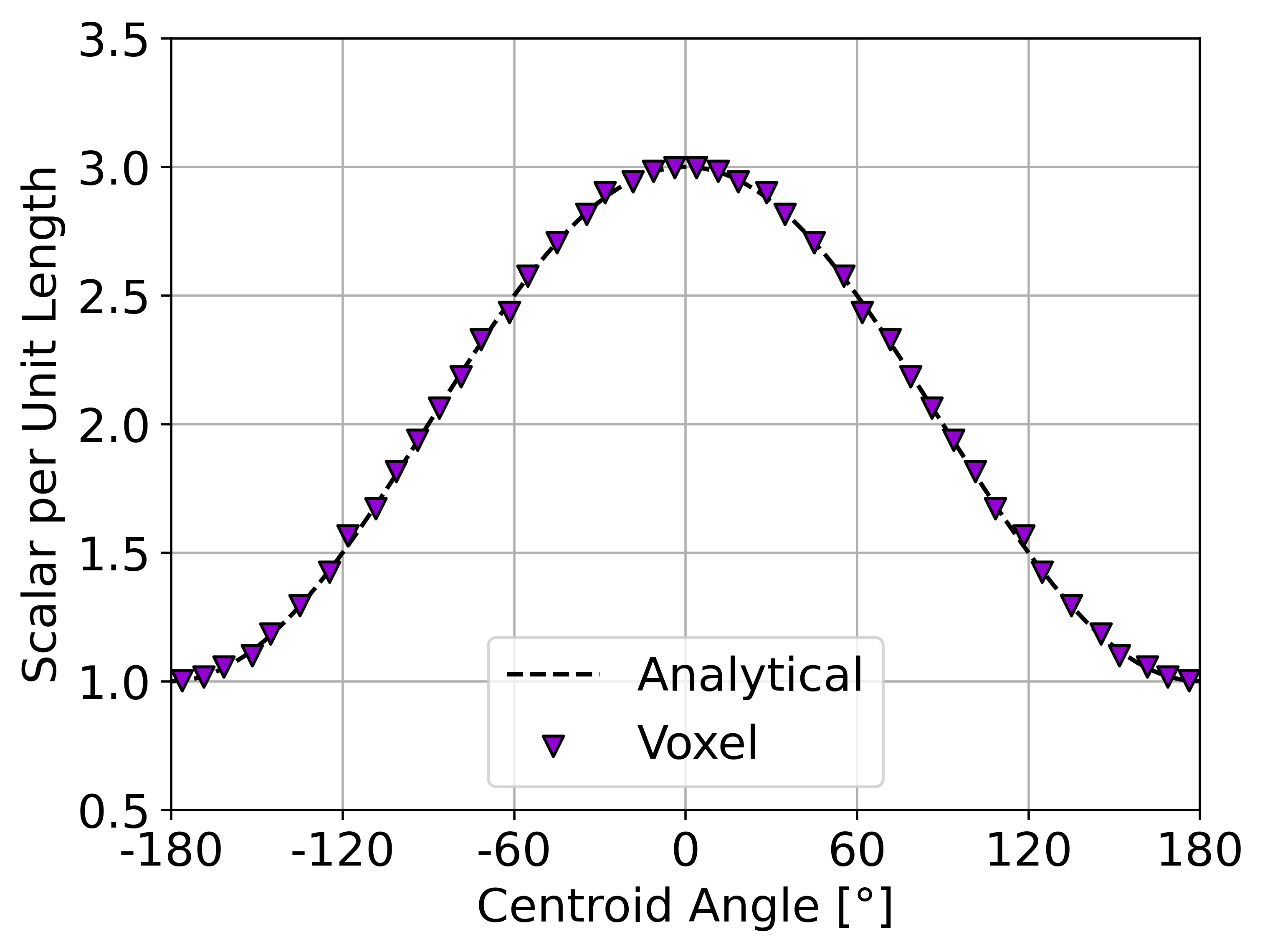}
    \caption{Triacontagon.}\label{fm_circle}
\end{subfigure}
\caption{Flux mapping applied to the convex geometries of Fig.~\ref{mm_geom}. A sinusoidal flux function (dashed line) was used to apply scalar values to the motion mapping surface elements. Flux mapping was then performed to divide the scalar quantities between the voxels (purple symbols).} 
\label{fm_purple}
\end{figure}

Results are given in Fig.~\ref{fm_purple} with both the analytical flux from Eq.~\ref{eq_gamma} and the voxel fluxes. Voxel fluxes are indicated with purple symbols in each graph according to the angular coordinate of the voxel centroid.
Across all shapes, most voxel fluxes fall near the analytical value. For each shape, the average absolute error of voxel flux with respect to analytical flux across the surface is less than 2.5\%. The square, diamond, and circle voxel fluxes in particular average at or below 1.0\% error. These three geometries also produced the smoothest boundary surfaces in the motion mapping illustrated in Fig.~\ref{mm_geom}.
The triangle and pentagon have more noticeable deviations in their voxel fluxes than the other three shapes. The larger deviations of the triangle are present over a wide range of angles while those of the pentagon are more localized near the rightmost and leftmost vertices. These vertices are in the same region where voxels penetrate the most beyond the boundary surface (Fig.~\ref{mm_pentagon}). Additionally, the centroid containment errors for $v_r=8$, $g_r=8$ in Fig.~\ref{mm_heat} were 0\% for the square, diamond, and circle. In contrast, the containment errors were 1.3\% for the pentagon and 2.4\% for the triangle.
These results seem to indicate that flux mapping errors result chiefly from error in the motion mapping rather than from the flux mapping itself.

Based on the results of Fig.~\ref{fm_purple}, the flux mapping module is found to be capable of accurately transferring spatially varying surface quantities from surface elements to nearby voxels.

\FloatBarrier
\section{Surface Recession}
\label{sec_results}

\begin{figure}[ht!]
\centering
\begin{subfigure}{0.32\textwidth}
    \centering
    \includegraphics[width=0.99\textwidth]{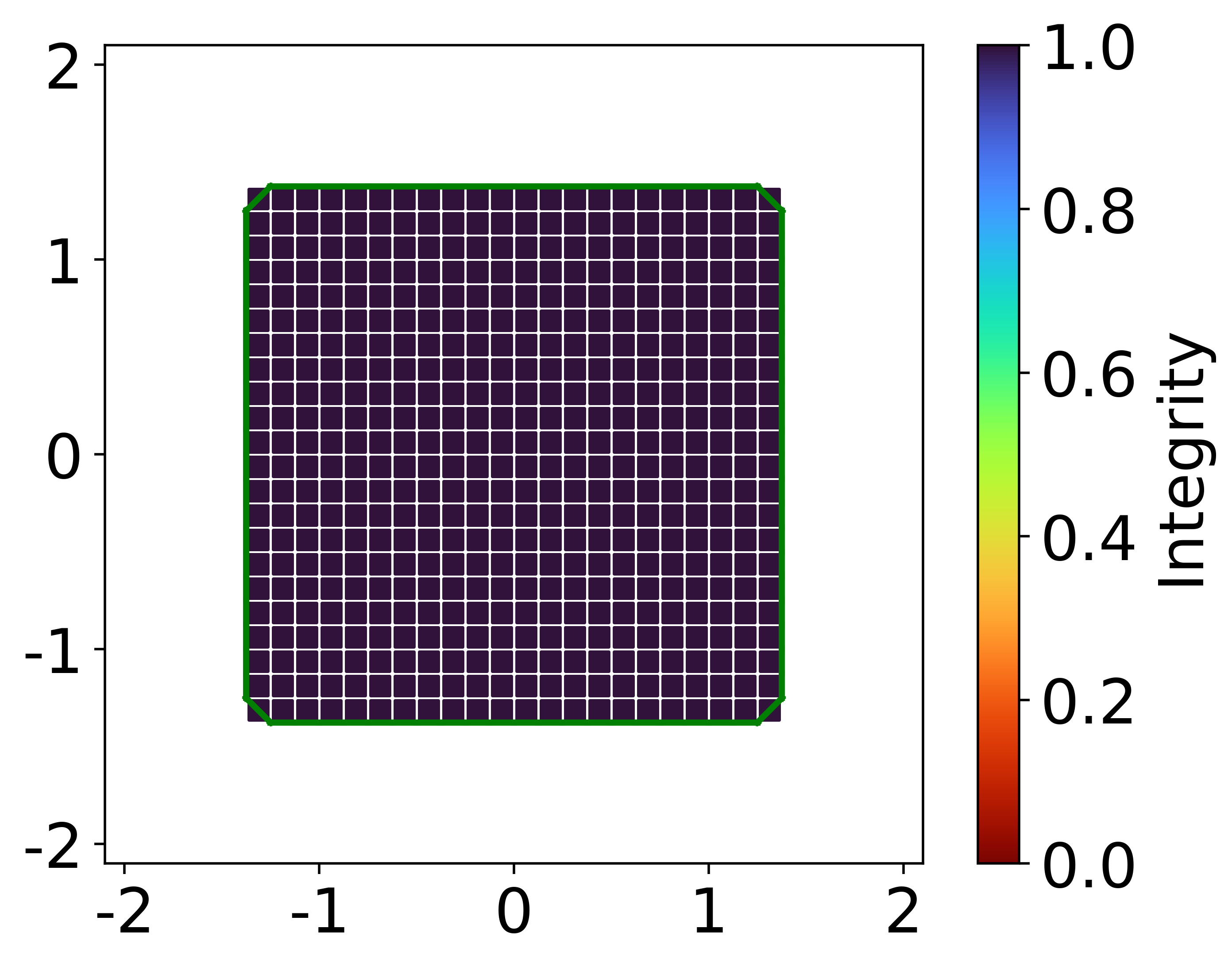}
    \caption{}\label{abl_square_first}
\end{subfigure}
\hfill
\begin{subfigure}{0.32\textwidth}
    \centering
    \includegraphics[width=0.99\textwidth]{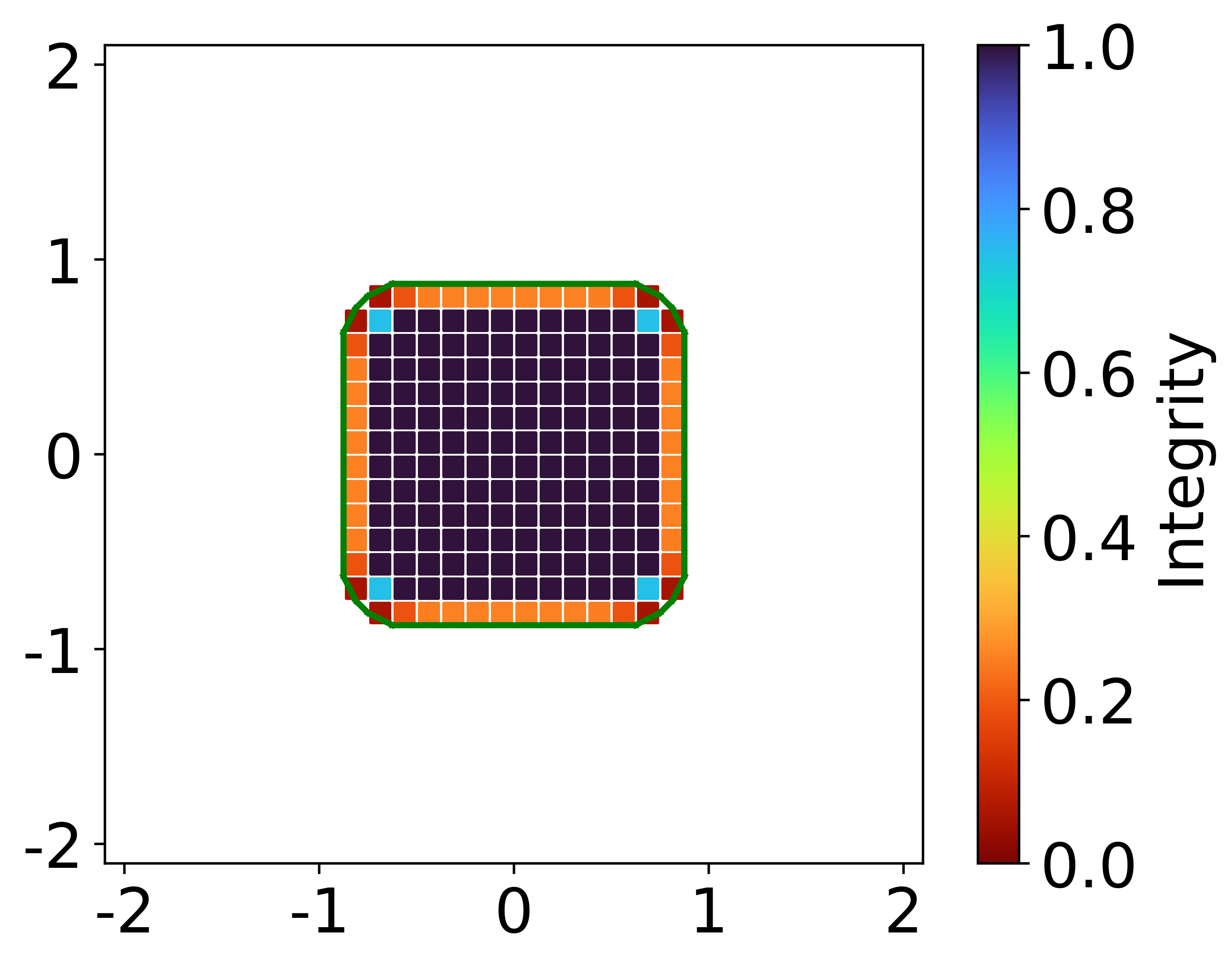}
    \caption{}\label{abl_square_mid}
\end{subfigure}
\hfill
\begin{subfigure}{0.32\textwidth}
    \centering
    \includegraphics[width=0.99\textwidth]{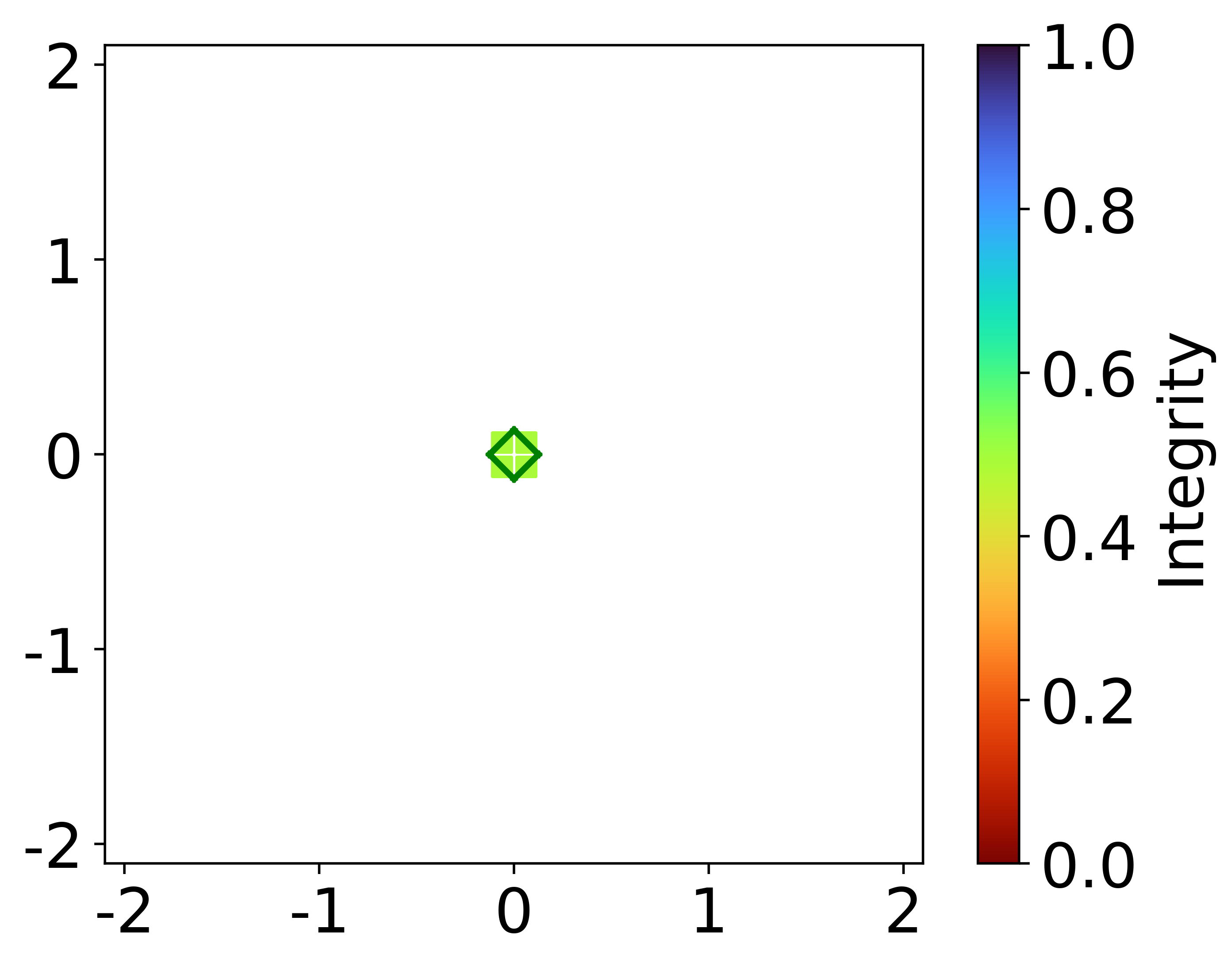}
    \caption{}\label{abl_square_last}
\end{subfigure}
\hfill
\begin{subfigure}{0.32\textwidth}
    \centering
    \includegraphics[width=0.99\textwidth]{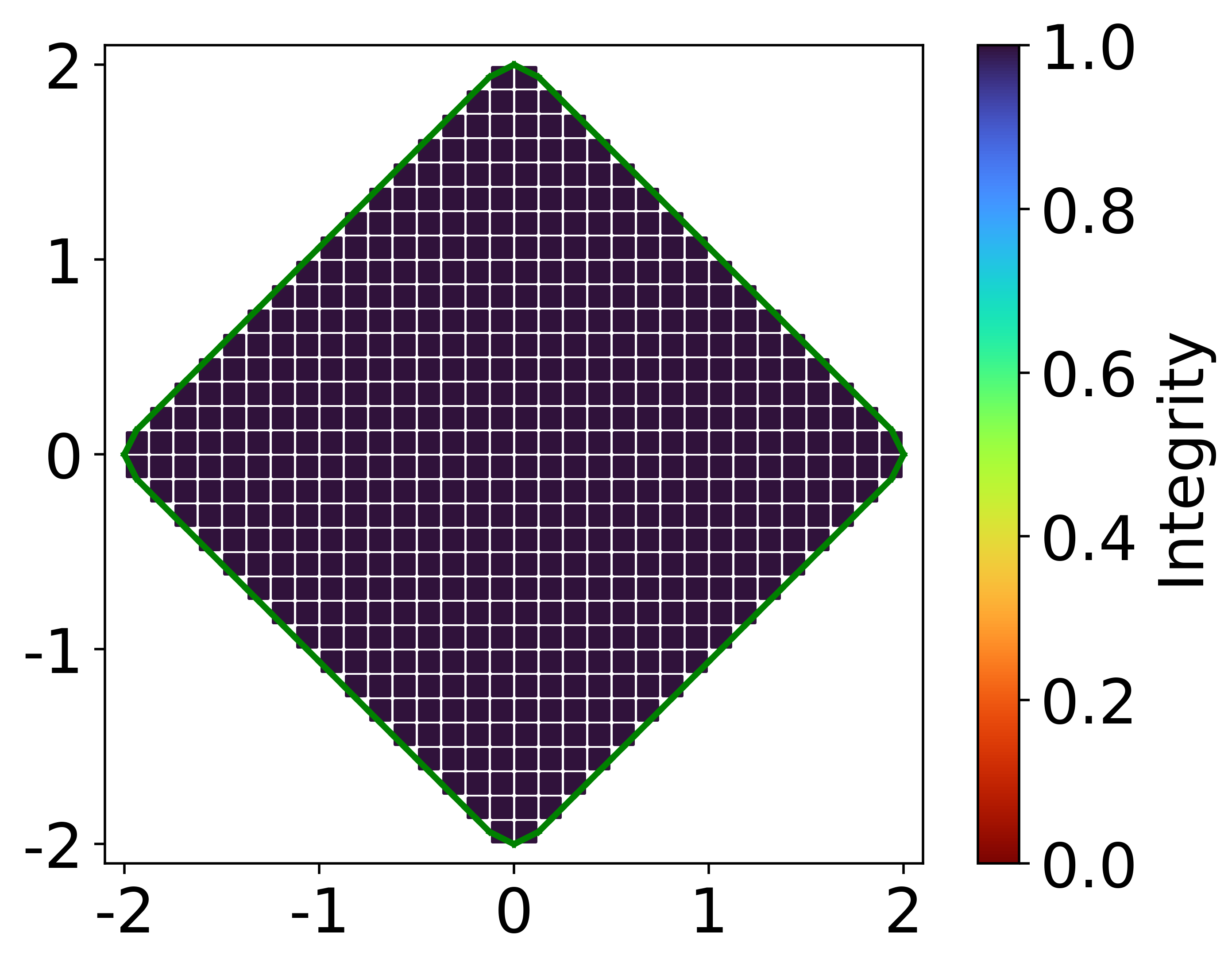}
    \caption{}\label{abl_diam_first}
\end{subfigure}
\hfill
\begin{subfigure}{0.32\textwidth}
    \centering
    \includegraphics[width=0.99\textwidth]{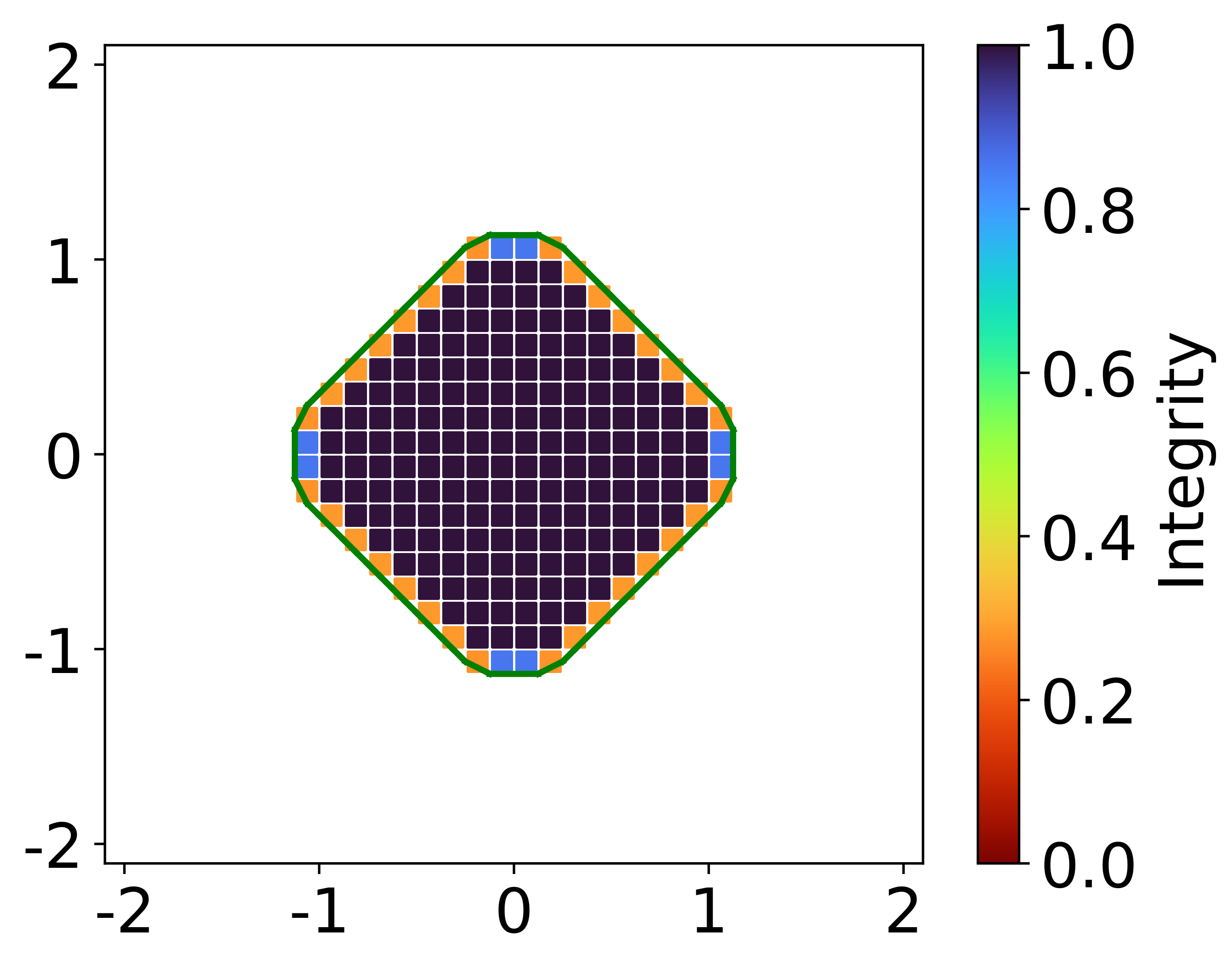}
    \caption{}\label{abl_diam_mid}
\end{subfigure}
\hfill
\begin{subfigure}{0.32\textwidth}
    \centering
    \includegraphics[width=0.99\textwidth]{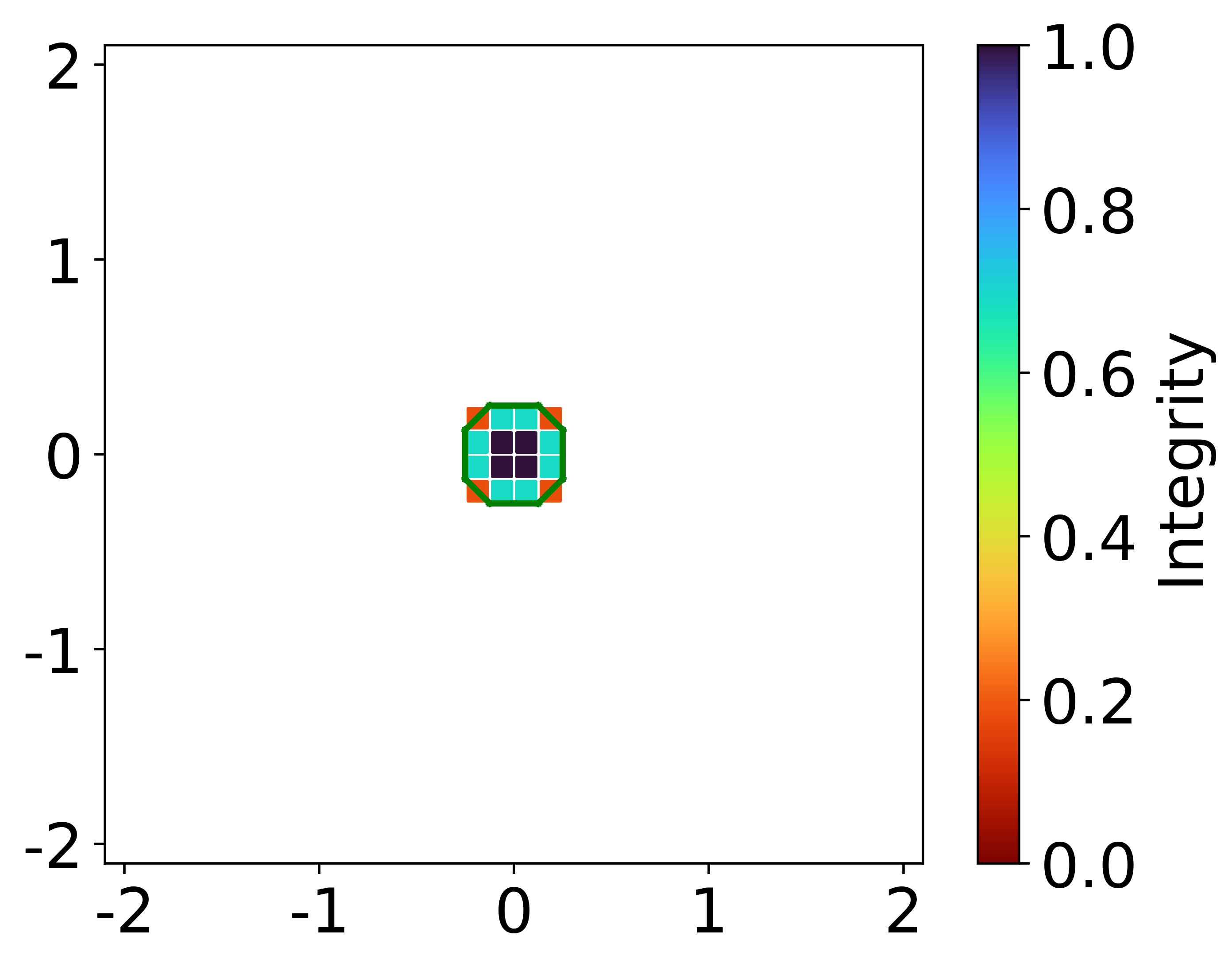}
    \caption{}\label{abl_diam_last}
\end{subfigure}
\hfill
\caption{Two polygonal geometries subjected to uniform, constant surface recession. First is a square before any recession (a), partway through the simulation (b), and towards the end of the simulation (c). A diamond geometry is plotted at the same three times (d-f). Each voxel is colored according to the fraction of material remaining in it. The green lines are the surfaces generated from motion mapping.} 
\label{abl_geom}
\end{figure}

In order to test the effectiveness of both marching windows modules working together in a single simulation, two fully coupled ablation simulations were performed. The square geometry and diamond geometry from Sec.~\ref{sec_motionmap} were simulated, as illustrated in Fig.~\ref{abl_geom}. For both simulations, voxel ratio and grid ratio were 16. A constant, uniform surface recession was applied to each surface element along its inward normal.

To simulate ablation in a loosely coupled manner with the given recession rate, each iteration of the simulation modeled the cycle of four stages illustrated in Fig.~\ref{fsi_coupling_scheme}. First, a surface is generated from the voxelized structure through motion mapping as described in Sec.~\ref{sec_motionmap}. The second stage is calculating recession rate ($\dot{d}$), defined as the ablated depth per time iteration. In a more complex ablation simulation, recession rate on each surface would be calculated with the help of a thermochemical fluid simulation technique. To test flux mapping accuracy, in the current work, a constant recession rate is set as
\begin{equation}
    \dot{d} = \frac{L_v}{4}
    \label{eq_recessrate}
\end{equation}
where $L_v$ is the side length of a voxel. Since the simulations are 2D, recession rate is equivalent to the volume loss rate per unit surface area, or volume flux. The third stage in each iteration is flux mapping. Voxels are assigned to surfaces by the flux mapping procedure described in Sec.~\ref{sec_fluxmap}, and the volume loss is distributed from surface elements to the voxels. The fourth and final stage is ablating each voxel according to its assigned volume loss. As a result of using a discrete timestep, the edge voxels may ablate beyond zero volume to negative volumes. To help prevent over-ablation of edge voxels, the recession rate in Eq.~\ref{eq_recessrate} was set such that the surface recedes less than one voxel length per timestep. Additionally, for any edge voxel that does over-ablate, the excess volume loss is transferred to neighboring voxels. The excess volume loss is distributed equally from the negative-volume edge voxel to the neighboring voxels of depth $d_v = 1$ that share a face. 

Using the settings described above, both simulations take approximately 40 iterations to run to completion with no remaining voxels. In Fig.~\ref{abl_geom}, each geometry is plotted before any ablation, during iteration 20, and during iteration 40. Each graphic illustrates the current remaining voxels and the marching squares surface produced from those voxels. The voxels are colored by their integrity, defined as the remaining volume fraction which has not ablated away. We note here that the integrity value does not have an effect on the weighting done during the motion mapping stage. If integrity goes to zero, then the voxel disappears and is weighted as a voxel void. In contrast, any voxel with integrity greater than zero is treated as a true voxel in the motion mapping stage.

Both the square and diamond voxelized geometries initially have tight-fitting surfaces (Figs.~\ref{abl_square_first},~\ref{abl_diam_first}). The round-off of the corners characteristic of marching squares can be seen on the square surface in Fig.~\ref{abl_square_first}, but only a small voxel area is cut off by this rounding. Both the square and diamond edges are smooth as the corresponding surfaces were in Fig.~\ref{mm_geom}. In the second image of each simulation (Figs.~\ref{abl_square_mid},~\ref{abl_diam_mid}), the surfaces are still tight-fitting to the voxels, but the ablation has rounded off the corners of the voxelized structures to some degree. In Fig.~\ref{abl_square_mid}, the ablation of internal voxels to prevent negative volumes in edge voxels can be seen at the corners. The light blue voxels at the corners are partially ablated internal voxels.

Surface deviations are only observed at the final stages of the transient evolution. The images from the late simulation in both cases (Figs.~\ref{abl_square_last},~\ref{abl_diam_last}) exemplify how the surface geometry can deviate from the voxelized structure as the total structure size approaches the size of a single marching windows cell. In Fig.~\ref{abl_square_last}, the remaining voxelized structure is a square, but the surface is a diamond shape. The voxelized structure in Fig.~\ref{abl_diam_last} is also square, but the surface is octagonal. While the surface deviations in Figs.~\ref{abl_square_last} and~\ref{abl_diam_last} may appear to be large compared to the remaining voxel structure, they are similar in size to the surface deviations present earlier in the simulation.

\begin{figure}[ht!]
\centering
\begin{subfigure}{0.49\textwidth}
    \centering
    \includegraphics[width=0.99\textwidth]{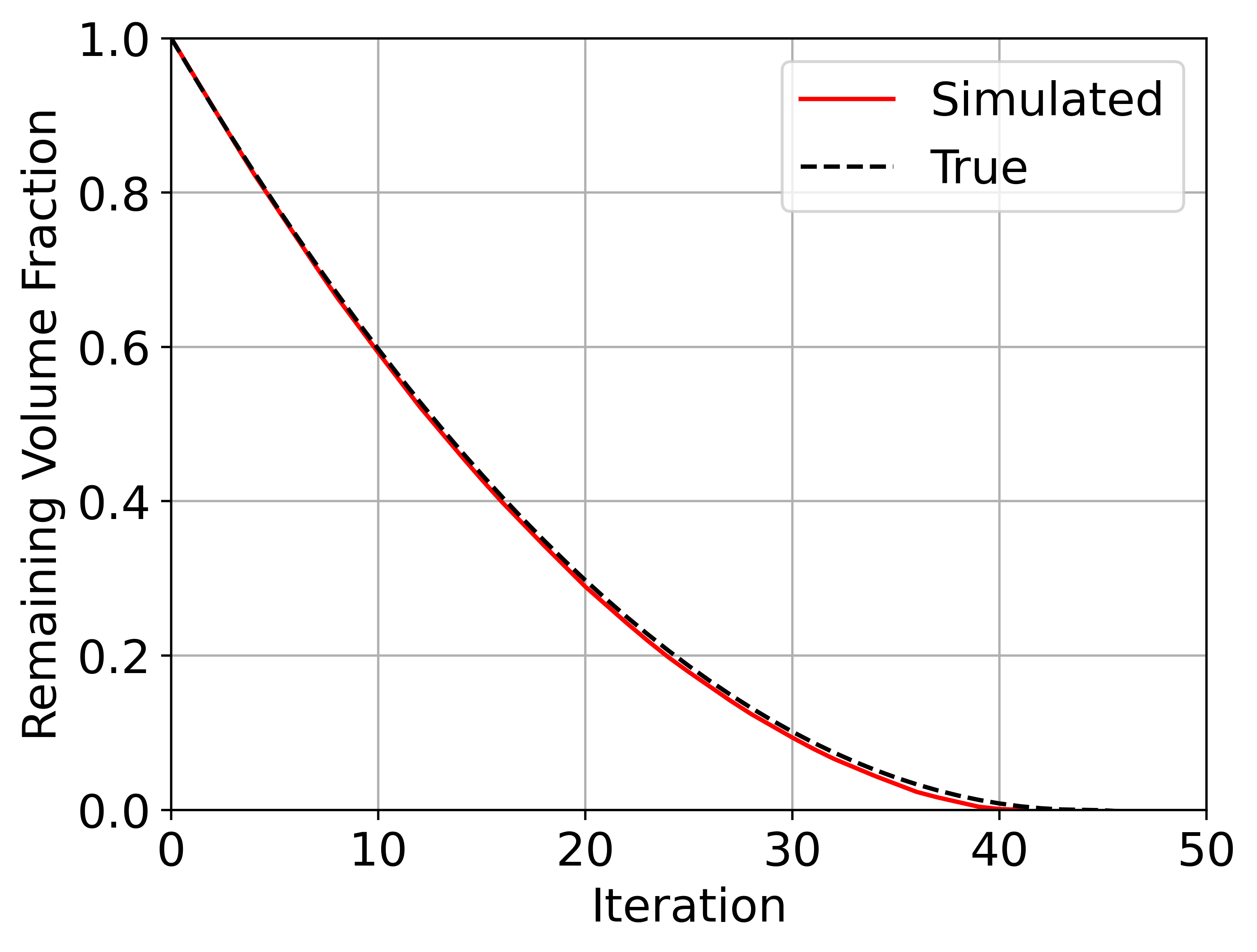}
    \caption{Square.}\label{abl_square_vol}
\end{subfigure}
\hfill
\begin{subfigure}{0.49\textwidth}
    \centering
    \includegraphics[width=0.99\textwidth]{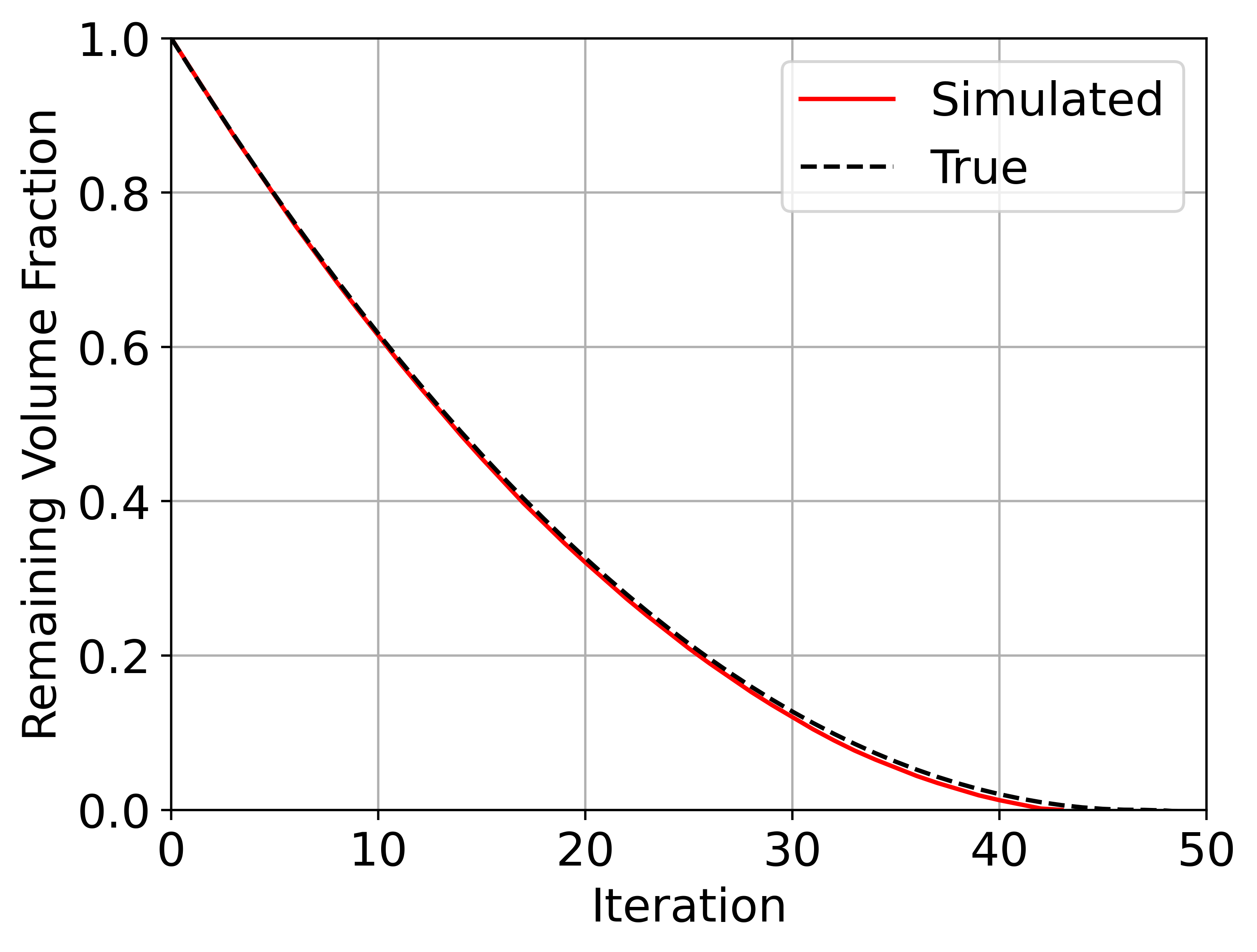}
    \caption{Diamond.}\label{abl_diam_vol}
\end{subfigure}
\hfill
\caption{Total volume over time for square (a) and diamond (b) ablation simulations. The analytical solution for a constant recession rate is given for each geometry alongside the simulation results for each iteration. Volumes are expressed as a fraction of the volume of the initial voxelized structure.} 
\label{abl_vol}
\end{figure}

The volume fractions of each geometry versus time are given in Fig.~\ref{abl_vol}. Since the square and diamond are congruent to each other, their true volume fraction evolutions can be expressed with the same function. At each time iteration ($\tau$) after ablation initiates, the true half-length of an edge ($d_\text{true}(\tau)$) can be expressed as
\begin{equation}
    d_\text{true}(\tau) = d(0) - \tau \dot{d}
\end{equation}
where $d(0)$ is the initial half-length of the solid. The current true volume with unit depth ($V_\text{true}(\tau)$) is
\begin{equation}
    V_\text{true}(\tau) = (2d_\text{true})^2
    \label{vd_eq}
\end{equation}
The simulated volume in both simulations is the sum of the volumes of all remaining voxels, weighted by their integrity. In Fig.~\ref{abl_vol}, the simulated and ideal volumes are both normalized by the original simulated volume ($V_\text{sim}(0)$). The original half-length used in Eq.~\ref{vd_eq} is determined based on the original simulated volume,
\begin{equation}
    d(0) = \frac{\sqrt{V_\text{sim}(0)}}{2}
    \label{d0_eq}
\end{equation}

In both the square and diamond simulations in Fig.~\ref{abl_vol}, the simulated volume loss nearly exactly matches the analytical solution. In both simulations, the ablation finishes more quickly than the true uniform ablation case, by 6.8\% for the square and by 7.8\% for the diamond. However, by the time the simulation is complete, less than 1\% of the analytical volume is left to ablate (0.5\% for the square and 0.6\% for the diamond), and the simulated volume fraction is very close to the analytical solution throughout both simulations.

The slight difference in the speed of ablation between the analytical and simulated solutions is likely a result of the discrete timestep. A constant depth recession rate is only approximated by the volume loss flux applied in these simulations. In using the volume loss approximation with a discrete timestep, the corners of the solid geometry lose volume from two sides simultaneously, resulting in overlapping recession that speeds up ablation relative to the analytical solution. This discretization effect can be mitigated by using a smaller timestep.







The results given in Fig.~\ref{abl_geom} and Fig.~\ref{abl_vol} indicate that marching windows is capable of coupling surfaces and voxelized structures. The receding geometries of Fig.~\ref{abl_geom} illustrate the capability of the motion mapping module to produce quality bounding surfaces. The simulated recession results of Fig.~\ref{abl_vol} similarly show that the flux mapping module can accurately transfer surface data to the underlying voxels.

\section{Conclusion}
\label{sec_conclusion}

Voxelized representations are often used to simulate solid structures in applications such as material ablation and mechanical response. However, voxel discretization does not provide a well-defined surface for multiphysics simulations that require external interactions with the solid structure. In the current work, a two-way coupling scheme, termed marching windows, was described. Marching windows bridges simulation domains defined by voxels with domains requiring boundary surfaces. The scheme consists of two related modules, one for motion mapping and the other for flux mapping. The motion mapping module takes a voxelized structure and marching squares grid parameters as input and generates a surface composed of connected line segments to approximate the boundary of the voxelized structure. The flux mapping module takes quantities calculated for these surface elements and transfers them to nearby voxels via geometric projections.
In this article, the motion mapping module was used on several geometries with varying voxel resolutions and grid cell resolutions to assess the fidelity of generated surfaces to the original voxelized structures. With mid-sized and small cell sizes, surface quality in all cases showed voxel containment errors below 2.5\%. Example geometries were also analyzed to evaluate the accuracy of the flux mapping module. Average flux transfer errors in all cases were below 2.5\% with respect to the original applied surface fluxes.
To assess the performance of both modules working together in a coupled framework, a constant-rate surface recession simulation was performed on two polygonal geometries using marching windows. The remaining simulated volume fraction closely tracked the analytical recession solution for the entirety of both simulations.
Future work for marching windows primarily involves analysis of 3D errors analogous to the calculations done in this work for 2D marching windows. Occlusion of voxel faces by other voxel faces may also need to be considered for general shapes. For the convex shapes considered in this article, the effect of occlusion was assumed to be minimal or nonexistent.



\section*{Data Availability}
Data will be made available on request.

\section*{Acknowledgments}
This material is based upon work supported by the Office of the Under Secretary of Defense for Research and Engineering under award number FA9550-22-1-0342. The views and conclusions contained in this document are those of the authors and should not be interpreted as representing the official policies, either expressed or implied, of the U.S. Government. The U.S. Government is authorized to reproduce and distribute reprints for Government purposes notwithstanding any copyright notation herein.

\FloatBarrier



\end{document}